\documentclass[preprintnumbers,aps,11pt,nofootinbib,tightenlines,floatfix]{revtex4}
\usepackage[paperwidth=210mm,paperheight=297mm,centering,hmargin=2.0cm,vmargin=2.925cm]{geometry}

\usepackage{amsmath,amssymb}
\usepackage{graphicx}
\usepackage{bm}
\usepackage{comment}
\usepackage{color}
\usepackage{subfigure}
\usepackage{array}
\usepackage{multirow}
\usepackage{natbib}
\usepackage{url}
\usepackage[]{units}

\usepackage[T1]{fontenc}
\usepackage[latin9]{inputenc}
\usepackage{graphicx}
\usepackage{esint}
\usepackage{hyperref}
\usepackage{atbegshi}
\usepackage{lipsum}
\usepackage{braket}

\begin{document}

\dimen\footins=5\baselineskip\relax

\preprint{\vbox{
\hbox{MIT-CTP-4688}
}}

\title{On the Implementation of General Background Electromagnetic Fields
\\
on a Periodic Hypercubic Lattice
}
\author{Zohreh Davoudi{\footnote{\tt davoudi@mit.edu}}
}
\affiliation{Center for Theoretical Physics, Massachusetts Institute of Technology, Cambridge, MA 02139, USA}

\author{William Detmold{\footnote{\tt wdetmold@mit.edu}}
}
\affiliation{Center for Theoretical Physics, Massachusetts Institute of Technology, Cambridge, MA 02139, USA}


\begin{abstract} 
	
Nonuniform background electromagnetic fields, once implemented in lattice quantum chromodynamics calculations of hadronic systems, provide a means to constrain a large class of electromagnetic properties of hadrons and nuclei, from their higher electromagnetic moments and charge radii to their electromagnetic form factors. We show how nonuniform fields can be constructed on a periodic hypercubic lattice under certain conditions and determine the precise form of the background $U(1)$ gauge links that must be imposed on the quantum chromodynamics gauge-field configurations to maintain periodicity. Once supplemented by a set of quantization conditions on the background-field parameters, this construction guarantees that no nonuniformity occurs in the hadronic correlation functions across the boundary of the lattice. The special cases of uniform electric and magnetic fields, a nonuniform electric field that varies linearly in one spatial coordinate (relevant to the determination of quadruple moment and charge radii), nonuniform electric and magnetic fields with given temporal and spatial dependences (relevant to the determination of nucleon spin polarizabilities) and plane-wave electromagnetic fields (relevant to the determination of electromagnetic form factors) are discussed explicitly.

\end{abstract}
\maketitle

\section{INTRODUCTION  
\label{sec:Intro} 
}
\noindent
The electromagnetic (EM) properties of hadrons and nuclei arise from the underlying interactions among their strongly interacting quark constituents and the photons. Consequently, any reliable theoretical  approach that aims to elucidate the electromagnetic structure of hadronic systems must necessarily account for the nonperturbative nature of quantum chromodynamics (QCD). Lattice QCD, which is a Monte Carlo evaluation of the QCD path integral regulated through a finite discrete spacetime, is the only such method by which to perform first-principles calculations of hadronic systems. It relies on the fundamental degrees of freedom of QCD, i.e., quarks and gluons, and as input, only takes the QCD parameters, the mass of quarks and the strength of the coupling constant. To be compared with nature, the results of lattice QCD calculations must be systematically extrapolated to the continuum and infinite-volume limits. Quantum electrodynamics (QED) can be treated along with QCD in a similar fashion, although the computational cost associated with a Monte Carlo evaluation of QCD and QED path integral is considerably higher, see Refs. \cite{Blum:2007cy,Basak:2008na,Blum:2010ym,Portelli:2010yn,Portelli:2012pn,Aoki:2012st,deDivitiis:2013xla,Borsanyi:2013lga,Drury:2013sfa, Borsanyi:2014jba, Blum:2014oka} for recent progress in this direction. Alternatively, given that EM interactions are perturbatively small, one can constrain the EM structure of hadrons and nuclei, for example their EM form factors, by evaluating the matrix elements of  current operators in the presence of solely QCD interactions. This approach has met with success in several cases, but is generally challenging (see Ref. \cite{Hagler:2009ni} for a review of lattice QCD calculations of nucleon structure). It is therefore useful to consider further alternatives.

When interested in the response of hadronic systems to weak EM fields, which determines properties such as EM moments and polarizabilities, a powerful method is to introduce electromagnetism through a classical background field. In this approach, fixed U(1) gauge links are simply imposed on the QCD gauge links in the lattice formulation. Depending on the computational resources available and the type of quantities that are being considered, this imposition may be performed solely on the valence quark sector of QCD -- where only the computation of quark propagators is influenced by the additional $U(1)$ links -- or on both the sea and valence quark sectors -- where the U(1) gauge links are also incorporated in the generation of the QCD gauge configurations. The former is not a reliable approximation when there are sea-quark (disconnected) contributions to hadronic correlation functions in background fields. A low-energy (multi-)hadronic theory that describes the interaction of the hadron (or nucleus) with weak external EM fields can be matched onto appropriate lattice QCD correlation functions. This matching constrains those parameters of the hadronic theory that characterize the response of the hadron (nucleus) to the applied external fields \cite{Lee:2013lxa, Lee:2014iha, davoudi2015, Caswell:1985ui, Labelle:1992hd, Labelle:1997uw, Kaplan:1998tg, Kaplan:1998we, Chen:1999tn, Detmold:2006vu, Detmold:2009dx, Hill:2012rh}. While uniform background fields already provide a means to constrain a wealth of EM quantities such as the magnetic moment and the electric and magnetic polarizabilities (see Refs. \cite{Bernard:1982yu, Martinelli:1982cb, Fiebig:1988en, Christensen:2004ca, Lee:2005ds, Lee:2005dq, Detmold:2006vu, Aubin:2008qp, Detmold:2009dx, Detmold:2010ts, Primer:2013pva, Lujan:2014kia, Beane:2014ora, Beane:2015yha, Chang:2015qxa} for recent progress in discerning EM properties of hadrons and light nuclei using lattice QCD with the background field method), more general background fields potentially provide sensitivity to many additional parameters of the low-energy theory. In particular, spin-dependent structure parameters beyond the magnetic moment can show up at low orders in the weak-field strength if one allows the spatial and/or temporal derivatives of the background fields to be nonvanishing. Among such quantities are spin polarizabilities of nucleons \cite{PhysRevD.47.3757, Detmold:2006vu}, and the quadrupole moment of hadrons and nuclei with spin, $s \geq 1$ \cite{davoudi2015}. Moreover, nonuniform background fields do not constrain one to the static limit of EM form factors, and by injecting energy and momentum into the system, provide a means to directly evaluate the corresponding off-forward hadronic matrix elements of the EM current from a response to the background fields \cite{Detmold:2004kw}, with different systematic uncertainties than other methods. Another application of such nonuniform background fields, as is recently proposed and implemented in Ref. \cite{Bali:2015msa}, is to evaluate the hadronic vacuum polarization function (as the leading hadronic contribution to the muon anomalous magnetic moment) through evaluating the magnetic susceptibilities with a plane-wave background field.

In order to explore such possibilities, one first needs to properly implement the desired EM background fields in lattice QCD calculations. A class of nonuniform EM fields, for example, have been implemented in Refs. \cite{Lee:2011gz, Engelhardt:2011qq} to obtain some preliminary results for spin polarizabilities of the nucleon \cite{Engelhardt:2011qq}. These studies do not consider background gauge potentials that are periodic at the boundary. Retaining the \emph{periodic boundary conditions} (PBCs) that are imposed on the gauge fields in the majority of lattice QCD calculations requires a nontrivial implementation of the $U(1)$ gauge links. However, such an implementation enforces smooth behavior of correlation functions across the boundary of the lattice and is a desired feature. Moreover, quantifying the behavior of the system in the hadronic theory is generally more straightforward in periodic EM potentials. Although nonperiodic implementations of uniform background fields have been pursued in some earlier lattice QCD studies (by placing hadronic sources away from the boundary effects) \cite{Bernard:1982yu, Martinelli:1982cb, Fiebig:1988en, Christensen:2004ca, Lee:2005ds, Lee:2005dq}, quantifying uncertainties associated with these nonuniformities is difficult \cite{Detmold:2008xk, Detmold:2009dx, Tiburzi:2013vza}. These issues can be prevented by explicitly modifying the naive $U(1)$ gauge links, and imposing conditions on the background field parameters, such that when setting the value of the gauge links at one boundary of the lattice to its value on the opposite boundary, no nonuniformities occur in the value of the $U(1)$ plaquettes. This guarantees smooth behavior of hadronic correlation functions across the boundaries. This paper presents in detail a procedure for the periodic implementation of background gauge fields, under certain conditions that are enumerated, along with several examples that follow from these general considerations.

A condition that supports periodic background $U(1)$ gauge fields on a hypercubic lattice\footnote{Although the words ``hypercubic'' and ``hypertorus'' are used throughout to refer to the lattice geometry, this paper considers the more general case of an anisotropic geometry where both the lattice spacing and the extent of the volume in temporal and spatial directions are different. In particular, the general result in Sec. \ref{sec:General-case} accounts for distinct lattice spacings and volume extents in all directions.} is well known for the case of uniform EM fields, namely the 't Hooft quantization condition (QC) \cite{'tHooft1979141, Smit:1986fn, AlHashimi:2008hr, Lee:2013lxa, Lee:2014iha}, and has been commonly implemented in lattice QCD calculations with the use of background fields \cite{Detmold:2006vu, Aubin:2008qp, Detmold:2009dx, Detmold:2010ts, Bali:2011qj, Primer:2013pva, Lujan:2014kia, Beane:2014ora, Beane:2015yha, Chang:2015qxa}. This condition requires the magnitude of the electric, $\mathbf{E}$, or magnetic, $\mathbf{B}$, field on a torus to be quantized, and follows from a simple argument: for a closed surface geometry, the net flux of the EM field through that surface is required to be quantized. The same condition can be obtained by imposing a more general boundary conditions, namely \emph{electro/magneto-PBCs}. These boundary conditions require the gauge and matter fields to be periodic up to a gauge transformation, and have been studied by 't Hooft for the case of Abelian and non-Abelian gauge theories, to explore the properties of the flux of the corresponding field strength tensors in the confinement regime  \cite{'tHooft19781,'tHooft1979141, 'tHooft1981}. When spacetime is discretized on a hypertorus, special care must be given to the implementation of the $U(1)$ gauge links at each spacetime points, see e.g., Refs. \cite{Damgaard:1988hh, Rubinstein:1995hc, Bali:2011qj}, to guarantee the values of the $U(1)$ elementary plaquettes remain constant in uniform background fields.

To implement background EM fields with arbitrary space-time dependences, a similar procedure must be undertaken to ensure that, despite the gauge links having been set to satisfy the PBCs at the boundary, the expected values of the $U(1)$ elementary plaquettes are correctly produced adjacent to the boundaries. This requirement enables one to determine the modified links near the boundary, as well as the conditions that the parameters of the chosen background fields must satisfy. We first demonstrate this procedure for a rather special case of an electric field generated from a scalar gauge potential with an arbitrary dependence on only one spatial coordinate. This allows one to generalize, rather straightforwardly, to the case of background fields generated from more general gauge potentials, but also makes one appreciate the subtleties and limitations encountered in the general case. We use this special case to study in detail the examples of a uniform electric field and an electric field that varies linearly in one spatial coordinate. We verify the periodicity of the setup by numerically evaluating the correlation functions of neutral pions in these background fields. This special case, along with the examples, is presented in Section \ref{sec:Special-case}. The general considerations are presented in Section \ref{sec:General-case}, where it is shown that a periodic implementation of gauge links in our framework is possible if the flux of the electric and magnetic fields through each plane on the lattice is independent of the coordinates transverse to that plane, allowing the flux to be quantized. The same condition also arises from consideration of the functions introduced in the modified links adjacent to the boundary such that the expected value of plaquettes are produced in all planes on the lattice. We take advantage of our general results, as summarized in Eqs. (\ref{eq:U-mu-general}-\ref{eq:QC-mu-nu-general}) of this paper, to work out several phenomenologically interesting examples. These examples focus on the background fields that give access to some of the spin polarizabilities of nucleons as suggested in Ref. \cite{Detmold:2006vu}, as well as the case of a plane-wave EM fields as proposed in Refs. \cite{Detmold:2004kw, Bali:2015msa}. These examples are presented in Section \ref{sec:Examples}. We summarize our results and conclude in Section \ref{sec:Conclusion}. An appendix is devoted to demonstrating the connection between our results concerning background $U(1)$ gauge fields with PBCs and those obtained under the imposition of certain electro/magneto-PBCs.

\section{A SPECIAL CASE: AN ELECTRIC FIELD ARISING FROM GAUGE POTENTIAL $A_{\mu} =\left(A_0(\mathbf{x}_3),\mathbf{0}\right)$
\label{sec:Special-case} 
}
\noindent
Let us choose a periodic $U(1)$ gauge field
\begin{eqnarray}
A_{\mu}\equiv (A_0,-\mathbf{A}) =\left(A_0(\mathbf{x}_3-\left[\frac{\mathbf{x}_3}{L}\right]L),\bm{0}\right),
\label{eq:A-NU-general}
\end{eqnarray}
with an arbitrary dependence on the third component of the position three-vector, $\mathbf{x}_3$, such that a periodic electric field is generated in the $\mathbf{x}_3$ direction,
\begin{eqnarray}
\mathbf{E}=-\bm{\nabla} A_0=E(\mathbf{x}_3-\left[\frac{\mathbf{x}_3}{L}\right]L) \hat{\mathbf{x}}_3.
\label{eq:E-uniform}
\end{eqnarray}
With the use of the floor function in the argument of the functions, the fields are made periodic in $\mathbf{x}_3$ with periodicity $L$. As is seen, we have adopted a mostly negative signature for the Minkowski metric. Throughout this paper, we take any  boldfaced letter to denote a three-vector. We leave the letters boldfaced even if they correspond to the components of a three-vector as to distinguish them from the components of a Minkowski four-vector, e.g., $\mathbf{x}_{i}=-x_i$ for $x_{\mu}=(t,-\mathbf{x})$.

The background field is implemented in a lattice QCD calculation by multiplying the QCD gauge links by the $U(1)$ gauge links through a direct product. Explicitly, for the choice in Eq. (\ref{eq:A-NU-general}),
\begin{eqnarray}
U^{(\text{QCD})}_{\mu}(x) \rightarrow U_{\mu}^{(\text{QCD})}(x) ~ \times ~ e^{ie\hat{Q} A_0(\mathbf{x}_3-\left[\frac{\mathbf{x}_3}{L}\right]L)a_t \times \delta_{\mu,0}},
\end{eqnarray}
where $\hat{Q}$ denotes the electric charge operator. Here, and in what follows, we define $a_s$ and $a_t$ to denote the lattice spacings along the spatial and temporal directions of the lattice, respectively. For a periodic lattice with spatial extent $L$ and temporal extent $T$, the value of the plaquette with $0 \leq \mathbf{x}_3 < L-a_s$ and $0 \leq t < T-a_t$ in the $0-3$ plane is
\begin{eqnarray}
\mathcal{P}_{(0,3)}(\mathbf{x}_3,t) &=& U_{0}^{(\text{QCD})}(\mathbf{x}_3,t) ~ e^{ie\hat{Q} A_0(\mathbf{x}_3)a_t}~
 U_{3}^{(\text{QCD})}(\mathbf{x}_3,t+a_t)
 \nonumber\\
& \times & U_{0}^{(\text{QCD})\dagger}(\mathbf{x}_3+a_s,t) e^{-ie\hat{Q} A_0(\mathbf{x}_3+a_s)a_t} U_{0}^{(\text{QCD})\dagger}(\mathbf{x}_3,t) 
\nonumber\\
&=& e^{ie\hat{Q}\left[A_0(\mathbf{x}_3)-A_0(\mathbf{x}_3+a_s)\right]a_t}~\mathcal{P}^{(\text{QCD})}_{(0,3)}(\mathbf{x}_3,t),
\label{eq:plq-03}
\end{eqnarray}
where we have left implicit the dependence on the $\mathbf{x}_1$ and $\mathbf{x}_2$ coordinates. Note that in the continuum,
\begin{eqnarray}
\Phi_{(0,3)}^{(E)}(\mathbf{x}_3) \equiv 
\int_{t}^{t+a_t}dt' \int_{\mathbf{x}_3}^{\mathbf{x}_3+a_s}\mathbf{E}_3(\mathbf{x}_3')d\mathbf{x}_3' = \left[A_0(\mathbf{x}_3)-A_0(\mathbf{x}_3+a_s)\right]a_t,
\label{eq:flux-elem-03}
\end{eqnarray}
is nothing but the total electric flux through the surface area of the elementary plaquette originated from the point $(\mathbf{x}_3,t)$ in the $0-3$ plane. Therefore, the desired value of the plaquette is obtained in Eq. (\ref{eq:plq-03}).

Since the lattice action depends upon links that originate from points $\mathbf{x}_i=L-1$ or $t=T-1$ and end at points $\mathbf{x}_i=L$ or $t=T$, one is required to specify the boundary conditions. By choosing PBCs, we demand that the $U(1)$ gauge link to be periodic  according to Eq. (\ref{eq:A-NU-general}). Then the value of the link that originates from $\mathbf{x}_3=L$ is set equal to its value at the origin. As a result, one must examine more carefully the value of the plaquettes located at $\mathbf{x}_3=L-a_s$,
\begin{eqnarray}
\mathcal{P}_{(0,3)}(L-a_s,t) &=& U_{0}^{(\text{QCD})}(L-a_s,t) ~ e^{ie\hat{Q} A_0(L-a_s)a_t} ~ U_{3}^{(\text{QCD})}(L-a_s,t+a_t)
\nonumber\\
& \times & U_{0}^{(\text{QCD})\dagger}(0,t) ~e^{-ie\hat{Q} A_0(0) a_t} ~ U_{3}^{(\text{QCD})\dagger}(L-a_s,t)
\nonumber\\
&=& e^{ie\hat{Q}\left[A_0(L-a_s)-\widetilde{A_0}(L)\right]a_t}~\mathcal{P}^{(\text{QCD})}_{(0,3)}(L-a_s,t) \times e^{ie\hat{Q}\left[\widetilde{A_0}(L)-A_0(0)\right]a_t},
\label{eq:plq-Lminus1}
\end{eqnarray}
where $\widetilde{A_0}$ is defined as the de-periodified counterpart of $A_0$,  $\widetilde{A_0} \equiv A_0((\mathbf{x}_3-\left[\frac{\mathbf{x}_3}{L}\right])+\left[\frac{\mathbf{x}_3}{L}\right])$. Then the first phase factor correctly accounts for the total electric flux through a plaquette located at $\mathbf{x}_3=L-a_s$, while the last phase factor must be eliminated. To achieve this, we are free to introduce an additional link in the $\mathbf{x}_3$ direction,
\begin{eqnarray}
U^{(\text{QCD})}_{\mu}(x) \rightarrow U_{\mu}^{(\text{QCD})}(x) ~ \times ~ e^{ie\hat{Q}a_t A_0(\mathbf{x}_3-\left[\frac{\mathbf{x}_3}{L}\right]L) \times \delta_{\mu,0}} ~ \times ~
e^{ie\hat{Q}\left[A_0(0)-\widetilde{A_0}(L)\right](t-\left[\frac{t}{T}\right]T)\times\delta_{\mu,3}\delta_{\mathbf{x}_3,L-a_s}},
\label{eq:link-0-3}
\end{eqnarray}
such that the last phase is canceled from Eq. (\ref{eq:plq-Lminus1}). Clearly, the additional link does not affect the value of the adjacent plaquettes.

Since the additional link introduced in Eq. (\ref{eq:link-0-3}) is $t$ dependent, and given that the gauge link is also required to be periodic with respect to the time variable, one must study the value of the plaquette located at $\mathbf{x}_3=L-a_s$ and $t=T-a_t$ more closely. For this plaquette, which is located at the far corner of the lattice,\footnote{Strictly speaking, there is no notion of a corner in a toroidal geometry. However, once an origin is specified for the coordinate system, we can choose to define the corner as the plaquette located at $\mathbf{x}_3=L-a_s$ and $t=T-a_t$.}
\begin{eqnarray}
&& \mathcal{P}_{(0,3)}(L-a_s,T-a_t) = U_{0}^{(\text{QCD})}(L-a_s,T-a_t) ~e^{ie\hat{Q} A_0(L-a_s) a_t} ~ U_{3}^{(\text{QCD})}(L-a_s,0)
\nonumber\\
&& ~ \qquad \qquad \qquad  \times U_{0}^{(\text{QCD})\dagger}(0,T-a_t) ~e^{-ie\hat{Q} A_0(0) a_t}~ U_{3}^{(\text{QCD})\dagger}(L-a_s,T-a_t) ~ e^{-ie\hat{Q}\left[A_0(0)-\widetilde{A_0}(L)\right](T-a_t)}
\nonumber\\
&& ~~~ \qquad \qquad \qquad \qquad = e^{ie\hat{Q}\left[A_0(L-a_s)-\widetilde{A_0}(L)\right]a_t}~\mathcal{P}^{(\text{QCD})}_{(0,3)}(L-a_s,T-a_t) \times e^{-ie\hat{Q}\left[A_0(0)-\widetilde{A_0}(L)\right]T}.
\label{eq:plq-last}
\end{eqnarray}
Although the first phase factor correctly accounts for the total flux through a plaquette located at $\mathbf{x}_3=L-a_s$, the last factor modifies the value of this plaquette away from the desired value. By demanding 
\begin{eqnarray}
e^{ie\hat{Q}\left[A_0(0)-\widetilde{A_0}(L)\right]T}=1,
\label{eq:net-flux}
\end{eqnarray}
the value of this plaquette is fixed to its desired value. Eq. (\ref{eq:net-flux}) places a QC on the net flux of the electric field through the $0-3$ plane on the lattice,
\begin{eqnarray}
\Phi_{(0,3)}^{(\mathbf{E}),\text{net}} \equiv \left[A_0(0)-\widetilde{A_0}(L)\right]T=e\hat{Q}\int_{0}^{T} dt \int_{0}^{L}\mathbf{E}_3(\mathbf{x}_3)d\mathbf{x}_3=2\pi n,
\label{eq:QC}
\end{eqnarray}
where $n \in \mathbb{Z}$. This QC could indeed be deduced a priori by recalling that the $0-3$ plane represents a closed surface area due to PBCs (the surface area of a torus), and the net flux through this surface must necessarily be quantized. Eqs. (\ref{eq:link-0-3}) and (\ref{eq:QC}) are the main results of this section. Two particular cases of a uniform electric field and a linearly varying electric field along the $\mathbf{x}_3$ direction can be readily worked out from this general result, and are presented in the following.

\subsection*{Example I: A constant electric field in the $\mathbf{x}_3$ direction
\label{sec:Special-case-E1} 
}
\noindent
An external periodic $U(1)$ gauge field
\begin{eqnarray}
A_{\mu}=\left(-E\times (\mathbf{x}_3-R-\left[\frac{\mathbf{x}_3}{L}\right]L),\bm{0}\right),
\label{eq:A-uniform}
\end{eqnarray}
with constant $E$ and $R$, gives rise to a uniform electric field in the $\mathbf{x}_3$ direction,
\begin{eqnarray}
\mathbf{E}=E \hat{\mathbf{x}}_3.
\label{eq:E-uniform}
\end{eqnarray}
%
%
\begin{figure}[t!]
\begin{center}  
\includegraphics[scale=0.465]{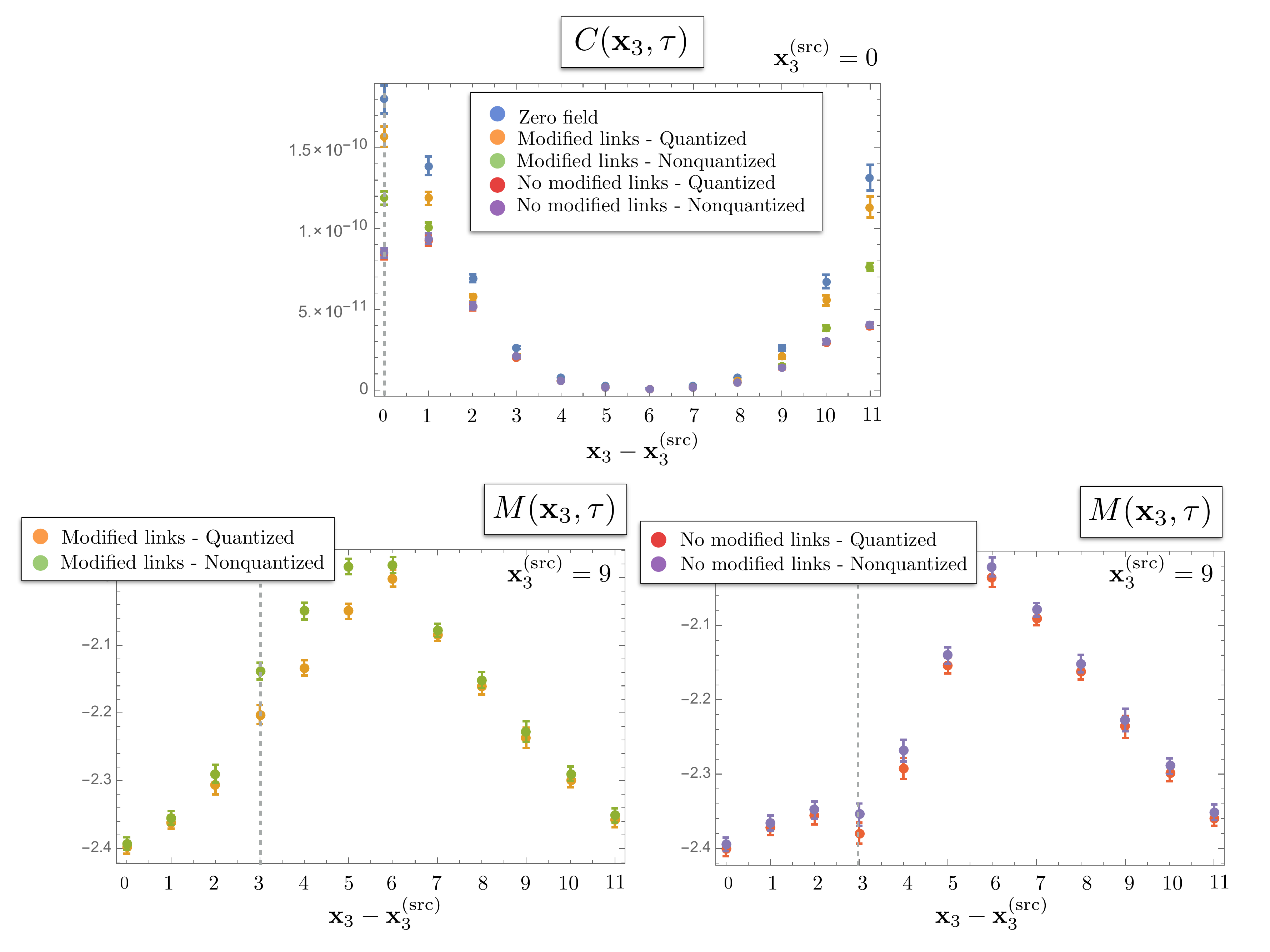}
\caption[.]{A comparison of choices of background gauge fields that result in a uniform electric field in the $\mathbf{x}_3$ direction: ``Modified links - Quantized'' denotes the choice of the gauge links in Eq. (\ref{eq:link-uniform}) with $R=0$, supplemented by the QC in Eq. (\ref{eq:QC-uniform}), with $n=3$. We have quantized the field for the d quark, corresponding to $|Q|=\frac{1}{3}$, which guarantees the quantization of the field for the u quark as well. ``Modified links - Nonquantized'' corresponds to the same choice of the gauge links but with a nonqunatized value of the field, $n=e$. ``No modified links - Quantized'' and ``No modified links - Nonquantized'' denote the naive choice of the gauge links without including the additional links in the $\mathbf{x}_3$ direction, with quantized and nonquantized values, $n=3$ and $n=e$, respectively. The upper panel shows the correlation function (projected to zero transverse momentum), $C(\mathbf{x}_3,\tau)$, as a function of $\mathbf{x}_3-\mathbf{x}_3^{\text{(src)}}$ at a fixed Euclidean time, $\tau/a_t=18$, while the lower panel depicts the dependence of the quantity $M(\mathbf{x}_3,\tau) \equiv \log \frac{C(\mathbf{x}_3,\tau)}{C(\mathbf{x}_3,\tau+1)}$ on $\mathbf{x}_3-\mathbf{x}_3^{\text{(src)}}$ at  $\tau/a_t=18$. In this demonstrative study, the lattice volume is taken to be $V=(12a_s)^3 \times (24a_t)$ with $a_s=a_t \approx 0.145 ~[\text{fm}]$, the QCD gauge configurations are quenched, the $U(1)$ gauge links are solely imposed on the QCD gauge links in the valence quark sector, and only the connected pieces of the correlation functions are evaluated for the neutral pion. The values of $\mathbf{x}_3$ and $\mathbf{x}_3^{(\text{src)}}$ in the figure are in units of $a_s$. The dashed lines denote the boundary of the lattice.}
\label{fig:pion-uniform}
\end{center}
\end{figure}
This electric field can be achieved by implementing the link,
\begin{eqnarray}
U^{(\text{QCD})}_{\mu}(x) \rightarrow U_{\mu}^{(\text{QCD})}(x) ~ \times ~ 
e^{-ie\hat{Q}Ea_t(\mathbf{x}_3-R-\left[\frac{\mathbf{x}_3}{L}\right]L) \times \delta_{\mu,0}} ~
e^{ie\hat{Q}EL(t-\left[\frac{t}{T}\right]T)\times\delta_{\mu,3}\delta_{\mathbf{x}_3,L-a_s}},
\label{eq:link-uniform}
\end{eqnarray}
which gives rise to the desired value of the plaquette in the $0-3$ plane
\begin{eqnarray}
\mathcal{P}_{(0,3)}(\mathbf{x}_3,t) = e^{ie\hat{Q}Ea_ta_s}~\mathcal{P}^{(\text{QCD})}_{(0,3)}(\mathbf{x}_3,t),
\label{eq:plq-x3}
\end{eqnarray}
for $0\leq \mathbf{x}_3 \leq L-a_s$ and $0\leq t \leq T-a_t$. In addition to the modification to the naive $U(1)$ links as presented in Eq. (\ref{eq:link-uniform}), the plaquettes at the boundary are correctly valued if the magnitude of the electric field is quantized,
\begin{eqnarray}
E=\frac{2\pi n}{e\hat{Q}TL},
\label{eq:QC-uniform}
\end{eqnarray}
with $n \in \mathbb{Z}$, in agreement with previous works, e.g., Refs. \cite{Smit:1986fn, Damgaard:1988hh, Rubinstein:1995hc, AlHashimi:2008hr, Lee:2013lxa, Lee:2014iha}.

We have numerically verified that, in contrast with the nonperiodic case where the additional link in the $\mathbf{x}_3$ direction is not implemented, the choice of the background gauge field in Eq. (\ref{eq:link-uniform}), supplemented by the QC in Eq. (\ref{eq:QC-uniform}), results in the periodicity of the correlation functions of neutral pions. The upper panel of Fig. \ref{fig:pion-uniform} shows the pion correlation function (projected to zero transverse momentum), $C(\mathbf{x}_3,\tau)$, as a function of $\mathbf{x}_3-\mathbf{x}_3^{\text{(src)}}$ at a fixed Euclidean time, $\tau/a_t=18$ ($\tau=it$). Explicitly, this correlation function is defined as
\begin{eqnarray}
C(\mathbf{x}_3,\tau) \equiv \left . C(\mathbf{x}_3,\tau;\mathbf{x}_3^{(\text{src})},0) \right|_{p_1=p_2=0} =
\sum_{\mathbf{x}_1=0}^{L-a_s}\sum_{\mathbf{x}_2=0}^{L-a_s}~ C(\mathbf{x},\tau;\mathbf{x}^{(\text{src})},0),
\label{eq:C-projected-def}
\end{eqnarray}
where
\begin{eqnarray}
C(\mathbf{x},\tau;\mathbf{x}^{(\text{src})},0)= \braket{0|\mathcal{O}_{\pi}(\mathbf{x},\tau)\mathcal{O}_{\pi}^{\dagger}(\mathbf{x}^{(\text{src})},0)|0}_{\mathbf{E}}.
\label{eq:C-def}
\end{eqnarray}
$\mathcal{O}_{\pi}^{\dagger}$($\mathcal{O}_{\pi}$) is a lattice interpolating operator that creates (annihilates) any hadronic states with the quantum numbers of the neutral pion. Subscript $\mathbf{E}$ refers to the fact that the expectation value is evaluated in the background of an electric field, $\mathbf{E}$. The calculation only involves imposing the $U(1)$ gauge links on the QCD gauge links in the valence sector. $\mathbf{x}^{\text{(src)}}$ denotes the location of the source, which for the upper panel is taken to be $\mathbf{x}^{(\text{src)}}=(0,0,0)$. Since for a neutral pion in a uniform electric field, the finite-volume correlation function with PBC must be symmetric about the point $\frac{L}{2}+\mathbf{x}_3^{\text{(src)}}$, the deviation of the correlation function from symmetricity for nonperiodic gauge-link choices, including those with the correct link structure but with nonquantized values of electric field, signals the breakdown of translational invariance in units of $L$ in the $\mathbf{x}_3$ direction (this translational invariance is the analogue of the magnetic translation group discussed in Ref. \cite{AlHashimi:2008hr} for a uniform magnetic field). Such breakdown is most evident in the quantity
\begin{eqnarray}
M(\mathbf{x}_3,\tau) \equiv \log \frac{C(\mathbf{x}_3,\tau)}{C(\mathbf{x}_3,\tau+1)},
\label{eq:M-def}
\end{eqnarray}
as is plotted as a function of $\mathbf{x}_3-\mathbf{x}_3^{\text{(src)}}$ for $\tau/a_t=18$ in the lower panel of Fig. \ref{fig:pion-uniform}. Here, the source is located $\mathbf{x}^{(\text{src)}}=(0,0,9a_s)$ and therefore the boundary point $\mathbf{x}_3=L \equiv 0$ corresponds to $\mathbf{x}_3-\mathbf{x}_3^{(\text{src)}}=3a_s$ in these plots. Nonuniformities in $M(\mathbf{x}_3,\tau)$ when crossing this boundary (denoted by the dashed line) are observed in all the cases considered, except for the ``Modified links - Quantized'' case.\footnote{These nonuniformities may be quantified more precisely by evaluating (the finite-difference approximation to) the derivative of the functions with respect to $\mathbf{x}_3$. As the continuum limit is approached, this (numerical) derivative diverges near the boundary as a result of nonperiodic implementations.} This is again a signature of losing translational invariance in units of $L$ in the $\mathbf{x}_3$ direction. In Refs. \cite{Detmold:2008xk, Detmold:2009dx}, a similar kinked feature was observed in the correlation function of neutral pions with nonperiodic implementations of a uniform electric field with the choice of a time-dependent gauge potential.

\subsection*{Example II: A linearly varying electric field in the $\mathbf{x}_3$ direction
\label{sec:Special-case-EII} 
}
\begin{figure}[t!]
\begin{center}  
\includegraphics[scale=0.425]{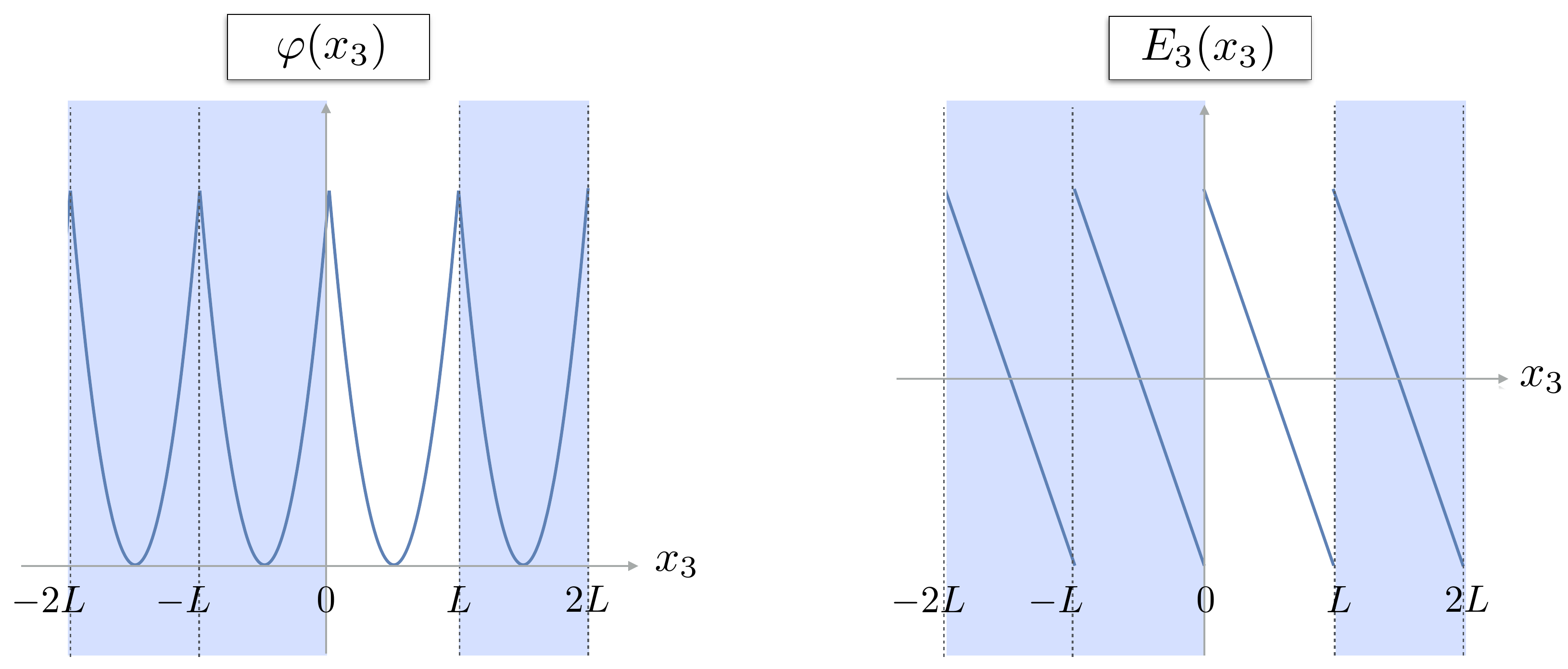}
\caption[.]{The scalar potential in Eq. (\ref{eq:A-nonuniform}) (the left panel) with $E_0<0$ and $R=\frac{L}{2}$ is a finite harmonic oscillator potential between $0 \leq \mathbf{x}_3 \leq L$. It produces a linearly varying electric field, Eq. (\ref{eq:E-nonuniform}), as depicted in the right panel. The periodic images of the potential and the electric field are also shown in the figures. 
}
\label{fig:NU-E}
\end{center}
\end{figure}

\noindent
An external periodic $U(1)$ gauge field
\begin{eqnarray}
A_{\mu}=\left(-\frac{E_0}{2} (\mathbf{x}_3-R-\left[\frac{\mathbf{x}_3}{L}\right]L)^2,\bm{0}\right),
\label{eq:A-nonuniform}
\end{eqnarray}
gives rise to a linearly varying electric field in the $\mathbf{x}_3$ direction,
\begin{eqnarray}
\mathbf{E}=E_0 \times (\mathbf{x}_3-R-\left[\frac{\mathbf{x}_3}{L}\right]L)\hat{\mathbf{x}}_3,
\label{eq:E-nonuniform}
\end{eqnarray}
as plotted in Fig. \ref{fig:NU-E}. 
This electric field can be implemented in a lattice QCD calculation through the following links,
\begin{eqnarray}
U^{(\text{QCD})}_{\mu}(x) \rightarrow U_{\mu}^{(\text{QCD})}(x) ~ \times ~ 
e^{-\frac{i}{2}e\hat{Q}E_0a_t(\mathbf{x}_3-R-\left[\frac{\mathbf{x}_3}{L}\right]L)^2 \times \delta_{\mu,0}} ~
e^{ie\hat{Q}E_0L(-R+\frac{L}{2})(t-\left[\frac{t}{T}\right]T)\times\delta_{\mu,3}\delta_{\mathbf{x}_3,L-a_s}},
\label{eq:link-nonuniform}
\end{eqnarray}
which gives rise to the following value for a plaquette in the $0-3$ plane
\begin{eqnarray}
\mathcal{P}_{(0,3)}(\mathbf{x}_3,t) = e^{ie\hat{Q}E_0a_ta_s(\mathbf{x}_3-R+\frac{a_s}{2})}~\mathcal{P}^{(\text{QCD})}_{(0,3)}(\mathbf{x}_3,t),
\label{eq:plq-x3}
\end{eqnarray}
for $0\leq \mathbf{x}_3 \leq L-a_s$ and $0\leq t \leq T-a_t$. There exists a QC when $R \neq {\frac{L}{2}}$ which constrains the value of the slope of the electric field to be quantized,
\begin{eqnarray}
E_0=\frac{2\pi n}{e\hat{Q}TL(-R+\frac{L}{2})},
\label{eq:QC-nonuniform}
\end{eqnarray}
with $n \in \mathbb{Z}$. When the offset value $R$ is chosen to be $\frac{L}{2}$, Eq. (\ref{eq:net-flux}) trivially holds and no quantization constraint is placed on $E_0$.\footnote{When there is no QC placed on the slope of the field, $E_0$, the field can become arbitrarily strong at the boundaries of the lattice in the $\mathbf{x}_3$ direction in the large-volume limit.  When interested in the response of the system to the external field at leading orders in the field strength, one needs to make sure that the field remains weak at any spacetime point on the lattice. This can be achieved by tuning $E_0$ to have appropriately small values, e.g., values that are power-law suppressed in volume.} Moreover, the additional link at point $\mathbf{x}_3=L-a_s$ in Eq. (\ref{eq:link-nonuniform}) will be equal to unity. 
%
\begin{figure}[t!]
\begin{center}  
\includegraphics[scale=0.4425]{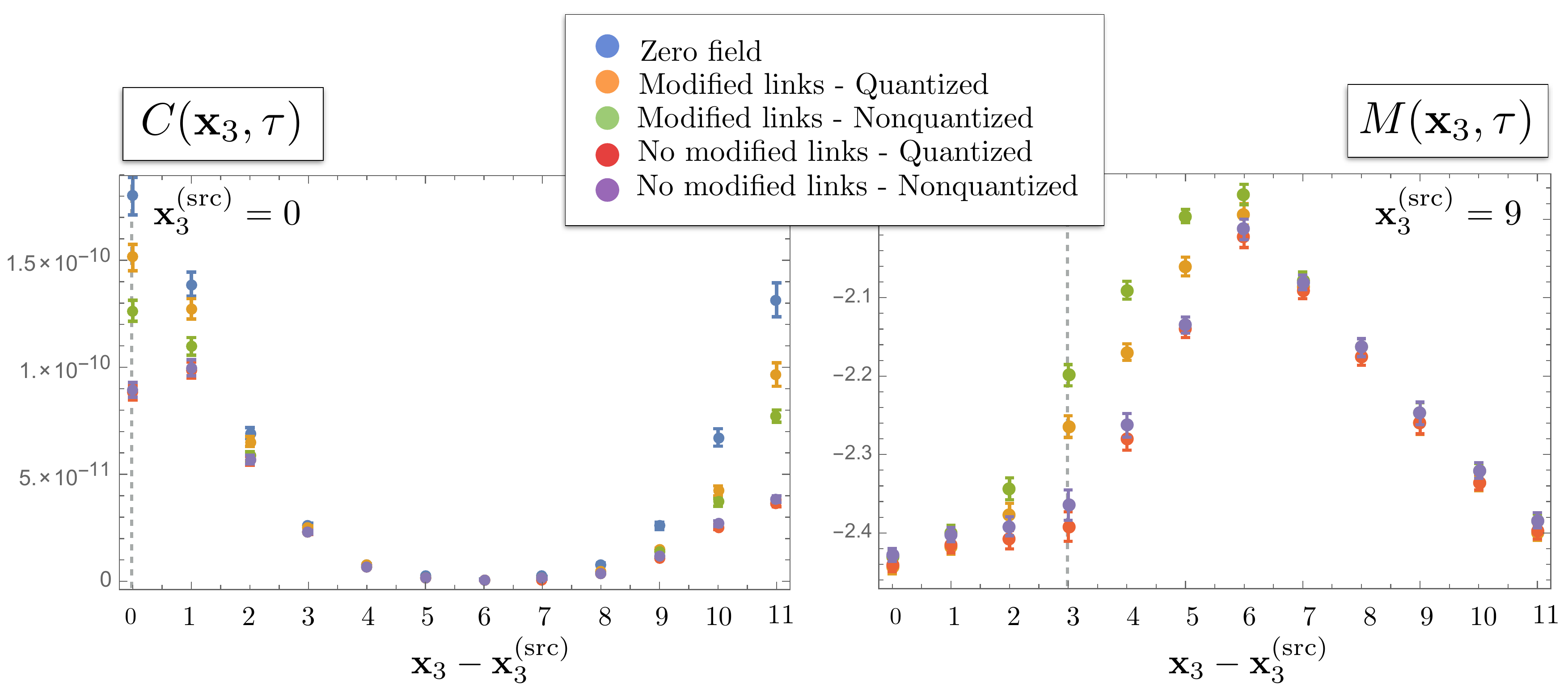}
\caption[.]{A comparison of choices of background gauge fields that result in a linearly varying electric field in the $\mathbf{x}_3$ direction: ``Modified links - Quantized'' denotes the choice of the gauge links in Eq. (\ref{eq:link-nonuniform}) with $R=0$, supplemented by the QC in Eq. (\ref{eq:QC-nonuniform}), with $n=3$ and $|Q|=\frac{1}{3}$, ``Modified links - Nonquantized'' corresponds to the same choice of the gauge links but with a nonqunatized value of the field slope, $n=e$. ``No modified links - Quantized'' and ``No modified links - Nonquantized'' denote the naive choice of the gauge links without including the additional links in the $\mathbf{x}_3$ direction, with quantized and nonqunatized values of the field slope, $n=3$ and $n=e$, respectively. The left panel shows the correlation function (projected to zero transverse momentum), $C(\mathbf{x}_3,\tau)$, as a function of $\mathbf{x}_3-\mathbf{x}_3^{\text{(src)}}$ at a fixed Euclidean time, $\tau/a_t=18$, while the left panel depicts the dependence of the quantity $M(\mathbf{x}_3,\tau) \equiv \log \frac{C(\mathbf{x}_3,\tau)}{C(\mathbf{x}_3,\tau+1)}$ on $\mathbf{x}_3-\mathbf{x}_3^{\text{(src)}}$ at  $\tau/a_t=18$. The details of this numerical study are the same as in Fig. \ref{fig:pion-uniform}. The values of $\mathbf{x}_3$ and $\mathbf{x}_3^{(\text{src)}}$ in the figure are in units of $a_s$. The dashed lines denote the boundary of the lattice.}
\label{fig:pion-nonuniform-R0}
\end{center}
\end{figure}

\begin{figure}[!]
\begin{center}  
\includegraphics[scale=0.4425]{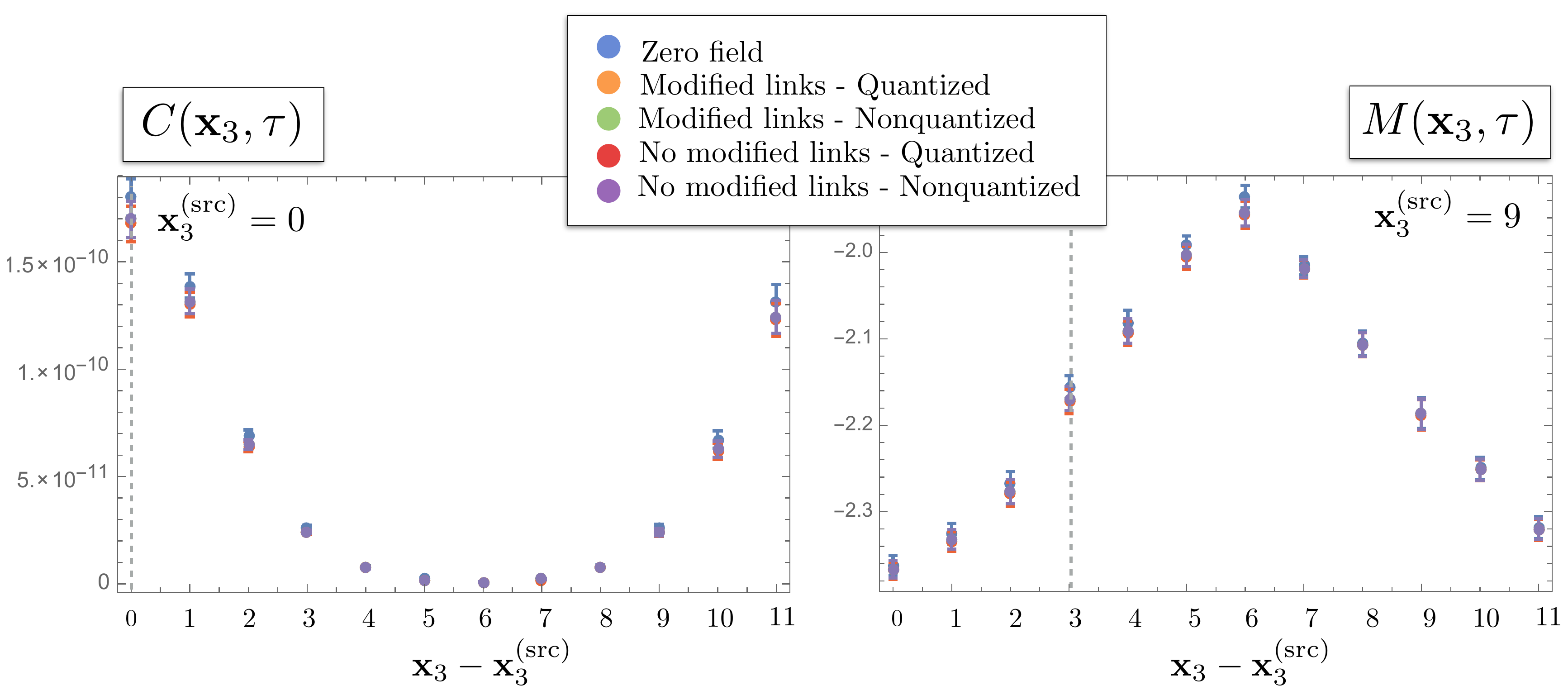}
\caption[.]{A comparison of choices of background gauge fields that result in a linearly varying electric field in the $\mathbf{x}_3$ direction, as described in the caption of Fig. \ref{fig:pion-nonuniform-R0}, but with the offset value being set to $R=\frac{L}{2}$. The left panel shows the correlation function (projected to zero transverse momentum), $C(\mathbf{x}_3,\tau)$, as a function of $\mathbf{x}_3-\mathbf{x}_3^{\text{(src)}}$ at a fixed Euclidean time, $\tau/a_t=18$, while the left panel depicts the dependence of the quantity $M(\mathbf{x}_3,\tau) \equiv \log \frac{C(\mathbf{x}_3,\tau)}{C(\mathbf{x}_3,\tau+1)}$ on $\mathbf{x}_3-\mathbf{x}_3^{\text{(src)}}$ at  $\tau/a_t=18$. The details of this numerical study are the same as in Fig. \ref{fig:pion-uniform}. The values of $\mathbf{x}_3$ and $\mathbf{x}_3^{(\text{src)}}$ in the figure are in units of $a_s$. The dashed lines denote the boundary of the lattice.}
\label{fig:pion-nonuniform-R6}
\end{center}
\end{figure}

We have implemented a linearly varying background electric field, with and without the additional links in the $\mathbf{x}_3$ direction, see Eq. (\ref{eq:link-nonuniform}), for quantized and nonqunatized values of the electric field slope, see Eq. (\ref{eq:QC-nonuniform}). Fig. \ref{fig:pion-nonuniform-R0} corresponds to the choices of electric field with the offset value $R=0$. Consequently, according to Eq. (\ref{eq:QC-nonuniform}), a QC is necessary to guarantee the full periodicity. In the left panel of the figure, the neutral pion correlation function (projected to zero transverse momentum), $C(\mathbf{x}_3,\tau)$, is plotted as a function of $\mathbf{x}_3-\mathbf{x}_3^{\text{(src)}}$ at a fixed Euclidean time, $\tau/a_t=18$. For a neutral pion in an electric field that varies as a function of $\mathbf{x}_3$, the finite-volume correlation function with PBCs will no longer be symmetric about the point $\frac{L}{2}+\mathbf{x}_3^{\text{(src)}}$ (see Ref. \cite{davoudi2015} and discussions associated with Fig. \ref{fig:pion-nonuniform-R6}) -- a feature that is indeed observed in the left panel of the figure. However, by displacing the source to $\mathbf{x}_3^{(\text{src})}\neq0$, the breakdown of translational invariance in units of $L$ is again manifest through nonuniformities when crossing the boundary of the lattice. This feature, as is plotted in the right panel of the figure, is observed in quantity $M(\mathbf{x}_3,\tau)$ at a fixed time when crossing $\mathbf{x}_3-\mathbf{x}_3^{(\text{src)}}=3a_s$, with $\mathbf{x}_3^{(\text{src})}=9a_s$, and is more prominent for implementations that do not use the modified values of the gauge links near the boundary.

%

According to Eq. (\ref{eq:link-nonuniform}), when the field offset value is set to $R=\frac{L}{2}$, no additional link is needed to guarantee periodicity. As a result, our implementations of both cases are trivially identical as is shown in Fig. \ref{fig:pion-nonuniform-R6}. However, a nontrivial check is to show that the periodicity of correlation functions is retained with arbitrary (nonqunatized) values of the electric field slope. Although there exists no QC for this case, we have continued to label the choices of the electric field slope as in the case of $R=0$. Then, as is evident from the right panel of Fig. \ref{fig:pion-nonuniform-R6}, the function $M(\mathbf{x}_3,\tau)$ smoothly crosses over the boundary of the lattice (up to the uncertainty of the data points), and correlation functions respect the periodicity of the fields. Additionally, the left panel of the figure suggests that the correlation function is approximately symmetric about the point $\frac{L}{2}+\mathbf{x}_3^{\text{(src)}}$ when $R=\frac{L}{2}$, in contrast with the case of $R=0$. This feature is understood by recalling that for a neutral pion, the leading dependence of the correlation function on a nonuniform external field arises from its charge radius.  The corresponding contribution then scales as $\sim E_0$ for the case of a linearly varying field in Eq. (\ref{eq:E-uniform}), which is constant. The next contribution results from the nonvanishing polarizability of the pion and scales as $\mathbf{E}^2$. Although this contribution is $\mathbf{x}_3$ dependent, it only depends on the square of the field, a quantity which is symmetric about $\mathbf{x}_3=\frac{L}{2}$ if the field offset value is set to $R=\frac{L}{2}$ (with $\mathbf{x}^{(\text{src})}_3=0$), see Fig. \ref{fig:NU-E}. This also explains the asymmetry in the neutral pion correlation function in the background of a linearly varying field with $R=0$, for which the field squared is not a symmetric function of $\mathbf{x}_3$ around the midpoint of the lattice in the $\mathbf{x}_3$ direction.

\section{A GENERAL CONSIDERATION: 
ELECTRIC AND MAGNETIC FIELDS ARISING FROM GAUGE POTENTIAL $A_{\mu} = \left(A_0(\mathbf{x},t),-\mathbf{A}(\mathbf{x},t)\right)$
\label{sec:General-case} 
}
\noindent
Having gained experience with the rather special case of an electric field generated from a scalar potential that only depended on a single space coordinate, we are ready to consider the case of a gauge potential with arbitrary dependences on all space and time coordinates. As will be demonstrated shortly, a periodic implementation of $U(1)$ gauge links can be achieved through modifying links adjacent to the boundary for a class of background fields that do not generate coordinate-dependent flux through any spacetime plane on the lattice (a closed surface on a torus).

In order to generate electric or magnetic fields with arbitrary dependences on space and time coordinates, one needs to consider a generic gauge potential with all four components being nonvanishing and having arbitrary space-time dependences,
\begin{eqnarray}
A_{\mu}(\mathbf{x},t)=\left(A_0(\mathbf{x},t),-\mathbf{A}(\mathbf{x},t)\right).
\label{eq:A-mu-general}
\end{eqnarray}
All the functions are to be understood to be periodified according to the floor-function prescription introduced in the previous section. Explicitly, all the $\mathbf{x}_i$ dependences of the functions must be understood as a dependence on $(\mathbf{x}_i-\left[\frac{\mathbf{x}_i}{L}\right]L)$. Similarly, all the time dependences are to be replaced with $(t-\left[\frac{t}{T}\right]T)$ dependences. As introduced above, once we de-periodify the functions in either space or time coordinates, we place a tilde over them. If $0 \leq \mathbf{x}_i < L$ and/or $0 \leq t < T$, there is no difference between functions with and without the tilde, and in the following this distinction only makes a difference when the functions are to be evaluated at $\mathbf{x}_i = L$ and/or $t = T$.

In the covariant formulation, the electric and magnetic fields are closely related and one does not need to distinguish the $0$ and $i$ components of vectors (where $i=1,2,3$ refers to spatial indices). However, to carefully account for different lattice spacings and volume extents in temporal and spatial directions, the plaquettes in the $0-i$ and $i-j$ planes, and consequently the cases of electric and magnetic fields, are studied separately. The goal is to generate the following electric and magnetic fields 
\begin{eqnarray}
\mathbf{E}(\mathbf{x},t)=-\frac{\partial \mathbf{A}(\mathbf{x},t)}{\partial t}-\bm{\nabla}A_0(\mathbf{x},t),
\nonumber\\
\mathbf{B}(\mathbf{x},t)=\bm{\nabla} \times \mathbf{A}(\mathbf{x},t),
\label{eq:E-B-general}
\end{eqnarray}
in the continuum limit by imposing the $U(1)$ gauge links on the QCD gauge links. Naively, one can start with the following links
\begin{eqnarray}
U^{(\text{QCD})}_0(\mathbf{x},t) & \rightarrow & U_0^{(\text{QCD})}(\mathbf{x},t) \times e^{ie\hat{Q}A_0(\mathbf{x},t)a_t},
\nonumber\\
U^{(\text{QCD})}_i(\mathbf{x},t) & \rightarrow & U_i^{(\text{QCD})}(\mathbf{x},t) \times e^{ie\hat{Q}A_i(\mathbf{x},t)a_s},
\label{eq:naive-links-general}
\end{eqnarray}
which then must be supplemented by additional $U(1)$ phase factors adjacent to the boundaries of the lattice to ensure periodicity of gauge-invariant quantities. To find these additional phase factors, it suffices to study the value of elementary plaquettes throughout the lattice. We first note that the naive implementation of the $U(1)$ gauge links gives the correct value of plaquettes everywhere except at $\mathbf{x}_i=L-a_s$ and $t=T-a_t$. Explicitly, for $0 \leq \mathbf{x}_i < L-a_s$ and $0 \leq t < T-a_t$, the value of plaquette in the $0-i$ plane reads 
\begin{eqnarray}
\mathcal{P}_{(0,i)}(\mathbf{x}_i,t) &=& U_0^{(\text{QCD})}(\mathbf{x}_i,t) e^{ie\hat{Q}A_0(\mathbf{x}_i,t)a_t}  U_i^{(\text{QCD})}(\mathbf{x}_i,t+a_t) e^{ie\hat{Q}A_i(\mathbf{x}_i,t+a_t)a_s}
\nonumber\\
&\times& U_0^{(\text{QCD})\dagger}(\mathbf{x}_i+a_s,t) e^{-ie\hat{Q}A_0(\mathbf{x}_i+a_s,t)a_t}  U_i^{(\text{QCD})\dagger}(\mathbf{x}_i,t) e^{-ie\hat{Q}A_i(\mathbf{x}_i,t)a_s}
\nonumber\\
&=& e^{-ie\hat{Q}\left[A_i(\mathbf{x}_i,t)-A_i(\mathbf{x}_i,t+a_t)\right]a_s+ie\hat{Q}\left[A_0(\mathbf{x}_i,t)-A_0(\mathbf{x}_i+a_s,t)\right]a_t} \times \mathcal{P}^{(\text{QCD})}_{(0,i)}(\mathbf{x}_i,t),
\label{eq:plq-xi-t}
\end{eqnarray}
where we have dropped the $\mathbf{x}_j$ and $\mathbf{x}_k$ dependences of the functions for brevity as they are fixed in this expression. Here, and in the following, $i,~j$ and $k$ assume distinct values. The $U(1)$ phase appearing in the value of the plaquette is clearly the expected value by noting that the electric flux through the corresponding surface area in the $0-i$ plane is
\begin{eqnarray}
&&\Phi_{(0,i)}^{(\mathbf{E})}(\mathbf{x}_j,\mathbf{x}_k)=\int_{\mathbf{x}_i}^{\mathbf{x}_i+a_s}d\mathbf{x}_i' \int_{t}^{t+a_t}dt' \mathbf{E}_i (\mathbf{x}_i',\mathbf{x}_j,\mathbf{x}_k,t')
\nonumber\\
&& ~~~ \qquad = -\left[ A_i(\mathbf{x}_i,\mathbf{x}_j,\mathbf{x}_k,t)-A_i(\mathbf{x}_i,\mathbf{x}_j,\mathbf{x}_k,t+a_t) \right]a_s+\left[ A_0(\mathbf{x}_i,\mathbf{x}_j,\mathbf{x}_k,t)-A_0(\mathbf{x}_i+a_s,\mathbf{x}_j,\mathbf{x}_k,t) \right]a_t
\nonumber\\
&& \qquad \qquad \qquad \qquad \qquad \qquad \qquad \qquad \qquad \qquad \qquad \qquad  \qquad \qquad \qquad \qquad +\mathcal{O}\left( a_s^2a_t,a_t^2a_s \right).
\label{eq:flux-as-at}
\end{eqnarray}
In the continuum limit, the flux per unit area, $\Phi_{(0,i)}^{(\mathbf{E})}/a_sa_t$, is finite and nonvanishing for any nonvanishing value of the electric field. As a result, the exponent of the $U(1)$ phase factor in Eq. (\ref{eq:plq-xi-t}) correctly accounts for the flux of electric field through the corresponding surface area in the continuum limit.

Since the value of the gauge links at the boundary will be set to their value at the origin due to the use of periodic functions in Eq. (\ref{eq:A-mu-general}), the value of the plaquettes adjacent to the boundary of the lattice will not necessarily produce their value as expected for the electric-field flux through these plaquettes. This can be easily seen by evaluating the value of plaquettes at $\mathbf{x}_i=L-a_s$ and $0 \leq t < T-a_t$,
\begin{eqnarray}
\mathcal{P}_{(0,i)}(\mathbf{x}_i=L-a_s,t) &=& U_0^{(\text{QCD})}(L-a_s,t) e^{ie\hat{Q}A_0(L-a_s,t)a_t}  U_i^{(\text{QCD})}(L-a_s,t+a_t) e^{ie\hat{Q}A_i(L-a_s,t+a_t)a_s}
\nonumber\\
& \times & U_0^{(\text{QCD})\dagger}(0,t) e^{-ie\hat{Q}A_0(0,t)a_t}  U_i^{(\text{QCD})\dagger}(L-a_s,t) e^{-ie\hat{Q}A_i(L-a_s,t)a_s}
\nonumber\\
& = & e^{-ie\hat{Q}\left[A_i(L-a_s,t)-A_i(L-a_s,t+a_t)\right]a_s+ie\hat{Q}\left[A_0(L-a_s,t)-\widetilde{A}_0(L,t)\right]a_t}\times\mathcal{P}_{(0,i)}^{(\text{QCD})}(L-a_s,t)
\nonumber\\
&& \qquad \qquad \qquad \qquad \qquad \qquad \qquad \qquad \qquad \qquad \times e^{ie\hat{Q}\left[\widetilde{A}_0(L,t)-A_0(0,t)\right]a_t},
\label{eq:plq-0i-Lminus1}
\end{eqnarray}
where again the $\mathbf{x}_j$ and $\mathbf{x}_k$ dependences of the functions are suppressed. One may now choose to modify the value of the link along either the $0$ or the $i$ directions in such a way to cancel out the extra phase in Eq. (\ref{eq:plq-0i-Lminus1}). However, it is easy to show that by modifying the link along the $0$ direction, the values of all the adjacent plaquettes are affected. In particular, one can not consistently make this modification throughout the lattice without shifting the value of another set of plaquettes from their desired values. However, one can  make the modification along the $i$ direction without changing the value of adjacent plaquettes. This is exactly the case we considered in the previous section, Eq. (\ref{eq:plq-Lminus1}) (upon setting $A_i=0$ in the current case). However, in contrast to the case considered previously, the scalar potential that appears in Eq. (\ref{eq:plq-0i-Lminus1}) carries an arbitrary $t$ dependence. As a result, the prescription of Eq. (\ref{eq:link-0-3}) will not work and we need to find a more general modified link in the $i$ direction to cancel the last phase factor in Eq. (\ref{eq:plq-0i-Lminus1}). It is easy to see that the following modification to the $U_i$ link at $\mathbf{x}_i=L-a_s$
\begin{eqnarray}
U_i^{(\text{QCD})}(\mathbf{x},t) \rightarrow U_i^{(\text{QCD})}(\mathbf{x},t) \times e^{ie\hat{Q}A_i(\mathbf{x},t)a_s} \times e^{ie\hat{Q}\left[A_0(\mathbf{x}_i=0,\mathbf{x}_j,\mathbf{x}_k,t)-\widetilde{A}_0(\mathbf{x}_i=L,\mathbf{x}_j,\mathbf{x}_k,t)\right] f_{i,0}(\mathbf{x}_j,\mathbf{x}_k,t) \times \delta_{\mathbf{x}_i,L-a_s}},
\nonumber\\
\label{eq:Ui-modified-general}
\end{eqnarray}
achieves the desired result provided that the boundary function $f_{i,0}(\mathbf{x}_j,\mathbf{x}_k,t)$ introduced above satisfies the following relation
\begin{eqnarray}
&&\left[A_0(\mathbf{x}_i=0,\mathbf{x}_j,\mathbf{x}_k,t+a_t)-\widetilde{A}_0(\mathbf{x}_i=L,\mathbf{x}_j,\mathbf{x}_k,t+a_t) \right]f_{i,0}(\mathbf{x}_j,\mathbf{x}_k,t+a_t)
\nonumber\\
&& ~~~ \qquad \qquad \qquad \qquad =\left[A_0(\mathbf{x}_i=0,\mathbf{x}_j,\mathbf{x}_k,t)-\widetilde{A}_0(\mathbf{x}_i=L,\mathbf{x}_j,\mathbf{x}_k,t)\right]\left(f_{i,0}(\mathbf{x}_j,\mathbf{x}_k,t)+a_t\right).
\label{eq:fi0-relation-I}
\end{eqnarray}
Note that $\mathbf{x}_j$ and $\mathbf{x}_k$ denote coordinates that are transverse to the $0-i$ plane. In order for the function $f_{i,0}$ to not spoil the periodicity of gauge links in the temporal and spatial directions, its dependences upon space and time coordinates must be understood through the floor-function prescription. Eq. (\ref{eq:fi0-relation-I}) is a recursive relation, and once the initial value of the function, corresponding to $f_{i,0}(\mathbf{x}_j,\mathbf{x}_k,t=0)$, is input, its value at every other point $t$, $f_{i,0}(\mathbf{x}_j,\mathbf{x}_k,t)$, can be obtained. For the special case of the previous section, $f_{i,0}$ is only a function of $t$ and satisfies a simple recursive relation, $f_{i,0}(t+a_t)=f_{i,0}(t)+a_t$. Once $f_{i,0}(0)$ is set to zero, the solution to this equation simply is $f_{i,0}(t)=t$, as already prescribed in Eq. (\ref{eq:link-0-3}). In general, if the function $f_{i,0}$ depends on $\mathbf{x}_j$ and $\mathbf{x}_k$ for the chosen gauge field, additional conditions must be placed on $f_{i,0}$. We will deduce these relations once we extend the above considerations to the plaquettes in other planes.

The $f_{i,0}$ function, as well as other gauge fields, have explicit time dependence and are periodic in this variable. As a result, fixing the value of plaquettes at $\mathbf{x}_i=L-a_s$ may not necessarily guarantee that its desired value is produced at $\mathbf{x}_i=L-a_s$ and $t=T-a_t$. However, before studying the value of this last plaquette, one needs to fix the value of all plaquettes located at $0 \leq \mathbf{x}_i < L-a_s$ and $t=T-a_t$. These plaquettes evaluate to
\begin{eqnarray}
&& \mathcal{P}_{(0,i)}(\mathbf{x}_i,t=T-a_t) = U_0^{(\text{QCD})}(\mathbf{x}_i,T-a_t) e^{ie\hat{Q}A_0(\mathbf{x}_i,T-a_t)a_t}  U_i^{(\text{QCD})}(\mathbf{x}_i,0) e^{ie\hat{Q}A_i(\mathbf{x}_i,0)a_s}
\nonumber\\
&& \qquad \qquad \qquad \times U_0^{(\text{QCD})\dagger}(\mathbf{x}_i+a_s,T-a_t) e^{-ie\hat{Q}A_0(\mathbf{x}_i+a_s,T-a_t)a_t}  U_i^{(\text{QCD})\dagger}(\mathbf{x}_i,T-a_t) e^{-ie\hat{Q}A_i(\mathbf{x}_i,T-a_t)a_s}
\nonumber\\
&& ~~~~~~ \qquad \qquad \qquad= e^{-ie\hat{Q}\left[A_i(\mathbf{x}_i,T-a_t)-\widetilde{A}_i(\mathbf{x}_i,T)\right]a_s} e^{ie\hat{Q}\left[A_0(\mathbf{x}_i,T-a_t)-A_0(\mathbf{x}_i+a_s,T-a_t)\right]a_t}\mathcal{P}_{(0,i)}^{(\text{QCD})}(\mathbf{x}_i,T-a_t)]
\nonumber\\
&& ~~ \qquad \qquad \qquad \qquad \qquad \qquad \qquad \qquad  \qquad \qquad \qquad \qquad \qquad \qquad \times e^{-ie\hat{Q}\left[\widetilde{A}_i(\mathbf{x}_i,T)-A_i(\mathbf{x}_i,0)\right]a_s},
\label{eq:plq-0i-Tminus1}
\end{eqnarray}
where we have suppressed the $\mathbf{x}_j$ and $\mathbf{x}_k$ dependences of the functions. To eliminate the unwanted phase factor, one can modify the value of the link along the $0$ direction as following
\begin{eqnarray}
U^{(\text{QCD})}_0(\mathbf{x},t) & \rightarrow & U_0^{(\text{QCD})}(\mathbf{x},t) \times e^{ie\hat{Q}A_0(\mathbf{x},t)a_t} \times e^{ie\hat{Q}\left[A_i(\mathbf{x},t=0)-\widetilde{A}_i(\mathbf{x},t=T)\right] f_{0,i}(\mathbf{x}) \times \delta_{t,T-a_t}},
\label{eq:U0-modified-general}
\end{eqnarray}
where the boundary function $f_{0,i}(\mathbf{x}_i)$ satisfies
\begin{eqnarray}
&& \left[A_i(\mathbf{x}_i+a_s,\mathbf{x}_j,\mathbf{x}_k,t=0)-\widetilde{A}_i(\mathbf{x}_i+a_s,\mathbf{x}_j,\mathbf{x}_k,t=T)\right]f_{0,i}(\mathbf{x}_i+a_s,\mathbf{x}_j,\mathbf{x}_k)
\nonumber\\
&& \qquad \qquad \qquad \qquad \qquad \qquad \qquad \qquad \qquad =\left[ A_i(\mathbf{x},t=0)-\widetilde{A}_i(\mathbf{x},t=T) \right]\left(f_{0,i}(\mathbf{x})+a_s\right).
\label{eq:f0i-relation-I}
\end{eqnarray}
The dependence of the new functions $f_{0,i}$ on any coordinate variable $\mathbf{x}_i$ must be realized through $(\mathbf{x}_i-\left[\frac{\mathbf{x}_i}{L}\right]L)$ as before. As will be shown shortly, although this condition on $f_{0,i}$ guarantees that the expected values of plaquettes at $0 \leq \mathbf{x}_i < L-a_s$ and $t=T-a_t$ are produced, it is not sufficient in general to ensure that the desired value of plaquettes is also produced in the $0-j$ plane with $j \neq i$ when $t=T-a_t$. We will return to this point below.

With the modifications of the $U_0$ and $U_i$ links according to Eqs. (\ref{eq:Ui-modified-general}) and (\ref{eq:U0-modified-general}), we are ready to inspect the value of the plaquette located at the far corner of the lattice with $\mathbf{x}_i=L-a_s$ and $t=T-a_t$. This plaquette evaluates to 
\begin{eqnarray}
&& \mathcal{P}_{(0,i)}(L-a_s,T-a_t) = U_0^{(\text{QCD})}(L-a_s,T-a_t) e^{ie\hat{Q}A_0(L-a_s,T-a_t)a_t}e^{ie\hat{Q}\left[A_i(L-a_s,0)-\widetilde{A}_i(L-a_s,T)\right]f_{0,i}(L-a_s)}
\nonumber\\
&& \qquad \qquad \qquad \qquad \times U_i^{(\text{QCD})}(L-a_s,0) e^{ie\hat{Q}A_i(L-a_s,0)a_s}e^{ie\hat{Q}\left[A_0(0,0)-\widetilde{A}_0(L,0)\right] f_{i,0}(0)}
\nonumber\\
&& \qquad \qquad \qquad \qquad \times U_0^{(\text{QCD})\dagger}(0,T-a_t) e^{-ie\hat{Q}A_0(0,T-a_t)a_t} e^{-ie\hat{Q}\left[A_i(0,0)-\widetilde{A}_i(0,T)\right]f_{0,i}(0)}
\nonumber\\
&& \qquad \qquad \qquad \qquad \times U_i^{(\text{QCD})\dagger}(L-a_s,T-a_t) e^{-ie\hat{Q}A_i(L-a_s,T-a_t)a_s}e^{-ie\hat{Q}\left[A_0(0,T-a_t)-\widetilde{A}_0(L,T-a_t)\right]f_{i,0}(T-a_t)}
\nonumber\\
&& \qquad \qquad \qquad \qquad = e^{-ie\hat{Q}\left[A_i(L-a_s,T-a_t)-\widetilde{A}_i(L-a_s,T)\right]a_s+ie\hat{Q}\left[A_0(L-a_s,T-a_t)-\widetilde{A}_0(L,T-a_t)\right]a_t} \times \mathcal{P}_{(0,i)}^{(\text{QCD})}(\mathbf{x}_i,t)
\nonumber\\
&& ~~~~ \qquad  \qquad \qquad \qquad \qquad \qquad \times \left[\prod_{\mathbf{x}_i=0}^{L-a_s} e^{ie\hat{Q}\left[ A_i(\mathbf{x}_i,0)-\widetilde{A}_i(\mathbf{x}_i,T) \right]a_s}\right]
\left[\prod_{t=0}^{T-a_t} e^{-ie\hat{Q}\left[ A_0(0,t)-\widetilde{A}_0(L,t) \right]a_t}\right],
\label{eq:plq-0i-Lminus1-Tminus1}
\end{eqnarray}
where we have used Eqs. (\ref{eq:fi0-relation-I}) and (\ref{eq:f0i-relation-I}) to arrive at the final expression. As is required, the value of the plaquette is independent of the $f_{i,0}$ and $f_{0,i}$ functions. In order for this plaquette to have the desired value, one can impose the condition
\begin{eqnarray}
\left[\prod_{\mathbf{x}_i=0}^{L-a_s} e^{-ie\hat{Q}\left[ A_i(\mathbf{x}_i,,\mathbf{x}_j,\mathbf{x}_k,t=0)-\widetilde{A}_i(\mathbf{x}_i,\mathbf{x}_j,\mathbf{x}_k,t=T) \right]a_s}\right]
\left[\prod_{t=0}^{T-a_t} e^{ie\hat{Q}\left[ A_0(\mathbf{x}_i=0,\mathbf{x}_j,\mathbf{x}_k,t)-\widetilde{A}_0(\mathbf{x}_i=L,\mathbf{x}_j,\mathbf{x}_k,t) \right]a_t}\right]=1,
\label{eq:QC-0i-general}
\end{eqnarray}
to set the extra phase factors in Eq. (\ref{eq:plq-0i-Lminus1-Tminus1}) to unity. This condition, and its implication for the allowed gauge field choices, require further discussion. Let us first point out that Eq. (\ref{eq:QC-0i-general}) arises from adding up the value of the $U(1)$ plaquettes that are corrected by introducing modified links throughout the lattice in the $0-i$ plane,
\begin{eqnarray}
&& \prod_{\mathbf{x}_i=0}^{L-a_s} \prod_{t=0}^{T-a_t}  \left[ e^{-ie\hat{Q}\left[A_i(\mathbf{x}_i,\mathbf{x}_j,\mathbf{x}_k,t)-A_i(\mathbf{x}_i,\mathbf{x}_j,\mathbf{x}_k,t+a_t)\right]a_s+ie\hat{Q}\left[A_0(\mathbf{x}_i,\mathbf{x}_j,\mathbf{x}_k,t)-A_0(\mathbf{x}_i+a_s,\mathbf{x}_j,\mathbf{x}_k,t)\right]a_t} \right]
\nonumber\\
&& ~~~ \qquad = \left[\prod_{\mathbf{x}_i=0}^{L-a_s} e^{-ie\hat{Q}\left[ A_i(\mathbf{x}_i,\mathbf{x}_j,\mathbf{x}_k,0)-\widetilde{A}_i(\mathbf{x}_i,\mathbf{x}_j,\mathbf{x}_k,T) \right]a_s}\right]
\left[\prod_{t=0}^{T-a_t} e^{ie\hat{Q}\left[ A_0(0,\mathbf{x}_j,\mathbf{x}_k,t)-\widetilde{A}_0(L,\mathbf{x}_j,\mathbf{x}_k,t) \right]a_t}\right]
\nonumber\\
&& ~~~ \qquad = e^{ie\hat{Q} \left[-\sum_{\mathbf{x}_i=0}^{L-a_s} \left[ A_i(\mathbf{x}_i,\mathbf{x}_j,\mathbf{x}_k,0)-\widetilde{A}_i(\mathbf{x}_i,\mathbf{x}_j,\mathbf{x}_k,T) \right]a_s+\sum_{t=0}^{T-a_t}\left[ A_0(0,\mathbf{x}_j,\mathbf{x}_k,t)-\widetilde{A}_0(L,\mathbf{x}_j,\mathbf{x}_k,t) \right]a_t \right]}.
\label{eq:sum-of-plaq-0i}
\end{eqnarray}
This latter exponent is $ie\hat{Q}$ times the net flux of electric field through the $0-i$ plane. Explicitly in the continuum limit,
\begin{eqnarray}
&& \Phi_{(0,i)}^{(\mathbf{E}),\text{net}}(\mathbf{x}_j,\mathbf{x}_k) = \int_{0}^{T}dt \int_{0}^{L}d\mathbf{x}_i ~ \mathbf{E}_i (\mathbf{x}_i,\mathbf{x}_j,\mathbf{x}_k,t)
\nonumber\\
&& ~ \qquad = -\int_{0}^{L}d\mathbf{x}_i \left[ A_i(\mathbf{x}_i,\mathbf{x}_j,\mathbf{x}_k,0)-\widetilde{A}_i(\mathbf{x}_i,\mathbf{x}_j,\mathbf{x}_k,T) \right]+
\int_{0}^{T}dt \left[ A_0(0,\mathbf{x}_j,\mathbf{x}_k,t)-\widetilde{A}_0(L,\mathbf{x}_j,\mathbf{x}_k,t) \right].
\nonumber\\
\end{eqnarray}
So the condition in Eq. (\ref{eq:QC-0i-general}) states that the net flux of the electric field  through $0-i$ plane must be quantized, $\Phi_{(0,i)}^{(\mathbf{E}),\text{net}}(\mathbf{x}_j,\mathbf{x}_k)=\frac{2\pi n}{e\hat{Q}}$ with $n \in \mathbb{Z}$. This condition can only hold if the left-hand side is independent of $\mathbf{x}_j$ and $\mathbf{x}_k$ coordinates. It therefore constrains the class of electric fields that can be implemented in this relatively general framework on a periodic lattice.\footnote{More general frameworks that allow more complicated space-time dependences for the fields may potentially exist.} As we will see below, the same conclusion can be drawn by inspecting the consistency of the conditions on the $f_{\mu,\nu}$ functions ($\mu,\nu=0,1,2,3$).

Let us not constrain our discussion to the classes of gauge fields that can be quantized through Eq. (\ref{eq:QC-0i-general}) for the moment, and continue to allow a general coordinate dependence for the gauge fields. We will obtain all such constraints that arise on the space-time dependence of gauge fields shortly. Having studied the case of electric field in the $i$ direction in great detail, it is easy to deduce the modified links as well as the QC for the scenario where electric field is nonvanishing along  other spatial directions. Firstly, Eq. (\ref{eq:U0-modified-general}) must be generalized as following to incorporate additional links in the $(j \neq i)$ directions,
\begin{eqnarray}
U^{(\text{QCD})}_0(\mathbf{x},t) & \rightarrow & U_0^{(\text{QCD})}(\mathbf{x},t) \times e^{ie\hat{Q}A_0(\mathbf{x},t)a_t} \times \prod_{i=1,2,3} e^{ie\hat{Q}\left[A_i(\mathbf{x},t=0)-\widetilde{A}_i(\mathbf{x},t=T)\right] f_{0,i}(\mathbf{x}) \times \delta_{t,T-a_t}}.
\label{eq:U0-modified-general-compl}
\end{eqnarray}
At a first glance, it might appear that a similar equation to that in Eq. (\ref{eq:Ui-modified-general}) must represent the value of the link in the $i$ direction with $i=1,2,3$. However, one should bear in mind that in this general scenario there exist a nonzero magnetic field and the value of the links in the $i$ direction can only be fully fixed after inspecting the value of plaquettes in all cartesian planes.

The value of a plaquette in the $i-j$ plane with $0 \leq \mathbf{x}_i < L-a_s$ and $0 \leq \mathbf{x}_j < L-a_s$ is
\begin{eqnarray}
\mathcal{P}_{(i,j)}(\mathbf{x}_i,\mathbf{x}_j) &=& U_i^{(\text{QCD})}(\mathbf{x}_i,\mathbf{x}_j) e^{ie\hat{Q}A_i(\mathbf{x}_i,\mathbf{x}_j)a_s}  U_j^{(\text{QCD})}(\mathbf{x}_i+a_s,\mathbf{x}_j) e^{ie\hat{Q}A_j(\mathbf{x}_i+a_s,\mathbf{x}_j)a_s}
\nonumber\\
& \times & U_i^{(\text{QCD})\dagger}(\mathbf{x}_i,\mathbf{x}_j+a_s) e^{-ie\hat{Q}A_i(\mathbf{x}_i,\mathbf{x}_j+a_s)a_s}  U_i^{(\text{QCD})\dagger}(\mathbf{x}_i,\mathbf{x}_j) e^{-ie\hat{Q}A_j(\mathbf{x}_i,\mathbf{x}_j)a_s}
\nonumber\\
&=& e^{ie\hat{Q}\left[A_i(\mathbf{x}_i,\mathbf{x}_j)-A_i(\mathbf{x}_i,\mathbf{x}_j+a_s)\right]a_s-ie\hat{Q}\left[A_j(\mathbf{x}_i,\mathbf{x}_j)-A_j(\mathbf{x}_i+a_s,\mathbf{x}_j)\right]a_s}~\mathcal{P}^{(\text{QCD})}_{(i,j)}(\mathbf{x}_i,\mathbf{x}_j),
\label{eq:plq-xi-xj-general}
\end{eqnarray}
which correctly accounts for the magnetic field flux through the surface area $a_s^2$ in the $i-j$ plane in the continuum limit,
\begin{eqnarray}
&& \Phi_{(i,j)}^{(\mathbf{B})}(\mathbf{x}_k,t) = -\frac{1}{2}\epsilon_{ijk} \int_{\mathbf{x}_i}^{\mathbf{x}_i+a_s}d\mathbf{x}_i' \int_{\mathbf{x}_j}^{\mathbf{x}_j+a_s}d\mathbf{x}_j' \mathbf{B}_k (\mathbf{x}_i',\mathbf{x}_j',\mathbf{x}_k,t)
\nonumber\\
&& ~ = \left[ A_i(\mathbf{x}_i,\mathbf{x}_j,\mathbf{x}_k,t)-A_i(\mathbf{x}_i,\mathbf{x}_j+a_s,\mathbf{x}_k,t) \right]a_s-\left[ A_j(\mathbf{x}_i,\mathbf{x}_j,\mathbf{x}_k,t)-A_j(\mathbf{x}_i+a_s,\mathbf{x}_j,\mathbf{x}_k,t) \right]a_s+\mathcal{O}\left( a_s^3 \right).
\nonumber\\
\label{eq:flux-as-as}
\end{eqnarray}
Eq. (\ref{eq:plq-xi-xj-general}) suggests that the results for the modified links, and the QC that was obtained from the consideration of the plaquettes in the $0-i$ plane, can be simply carried over to the case of the $i-j$ plane by substituting $0 \rightarrow i$, $i \rightarrow j$, $a_t \rightarrow a_s$ and $T \rightarrow L$ in those relations. In particular, one finds that 
\begin{eqnarray}
U^{(\text{QCD})}_i(\mathbf{x}) & \rightarrow & U_i^{(\text{QCD})}(\mathbf{x}) \times e^{ie\hat{Q}A_i(\mathbf{x},t)a_s} \times
e^{ie\hat{Q}\left[A_0(\mathbf{x}_i=0,\mathbf{x}_j,\mathbf{x}_k,t)-\widetilde{A}_0(\mathbf{x}_i=L,\mathbf{x}_j,\mathbf{x}_k,t)\right] f_{i,0}(\mathbf{x}_j,\mathbf{x}_k,t) \times \delta_{\mathbf{x}_i,L-a_s}}
\nonumber\\
&& \qquad \qquad \qquad \qquad  ~~ \times ~ \prod_{j \neq i}  e^{ie\hat{Q}\left[A_j(\mathbf{x}_i=0,\mathbf{x}_j,\mathbf{x}_k,t)-\widetilde{A}_j(\mathbf{x}_i=L,\mathbf{x}_j,\mathbf{x}_k,t)\right] f_{i,j}(\mathbf{x}_j,\mathbf{x}_k,t) \times \delta_{\mathbf{x}_i,L-a_s}},
\label{eq:Ui-modified-general-compl}
\end{eqnarray}
where the boundary function $f_{i,j}(\mathbf{x}_j)$ must satisfy
\begin{eqnarray}
&& \left[ A_j(\mathbf{x}_i=0,\mathbf{x}_j+a_s,\mathbf{x}_k,t)-\widetilde{A}_j(\mathbf{x}_i=L,\mathbf{x}_j+a_s,\mathbf{x}_k,t) \right]f_{i,j}(\mathbf{x}_j+a_s,\mathbf{x}_k,t)=
\nonumber\\
&& ~~~~ \qquad \qquad \qquad \qquad \left[ A_j(\mathbf{x}_i=0,\mathbf{x}_j,\mathbf{x}_k,t)-\widetilde{A}_j(\mathbf{x}_i=L,\mathbf{x}_j,\mathbf{x}_k,t) \right]\left(f_{i,j}(\mathbf{x}_j,\mathbf{x}_k,t)+a_s\right),
\label{eq:fij-relation-I}
\end{eqnarray}
\begin{eqnarray}
&& \left[ A_j(\mathbf{x}_i=0,\mathbf{x}_j,\mathbf{x}_k+a_s,t)-\widetilde{A}_j(\mathbf{x}_i=L,\mathbf{x}_j,\mathbf{x}_k+a_s,t) \right]f_{i,j}(\mathbf{x}_j,\mathbf{x}_k+a_s,t)=
\nonumber\\
&& \qquad \qquad \qquad \qquad \qquad \qquad \left[ A_j(\mathbf{x}_i=0,\mathbf{x}_j,\mathbf{x}_k,t)-\widetilde{A}_j(\mathbf{x}_i=L,\mathbf{x}_j,\mathbf{x}_k,t) \right] f_{i,j}(\mathbf{x}_j,\mathbf{x}_k,t),
\label{eq:fij-relation-II}
\end{eqnarray}
\begin{eqnarray}
&& \left[ A_j(\mathbf{x}_i=0,\mathbf{x}_j,\mathbf{x}_k,t+a_t)-\widetilde{A}_j(\mathbf{x}_i=L,\mathbf{x}_j,\mathbf{x}_k,t+a_t) \right]f_{i,j}(\mathbf{x}_j,\mathbf{x}_k,t+a_t)=
\nonumber\\
&& \qquad \qquad \qquad \qquad \qquad \qquad \left[ A_j(\mathbf{x}_i=0,\mathbf{x}_j,\mathbf{x}_k,t)-\widetilde{A}_j(\mathbf{x}_i=L,\mathbf{x}_j,\mathbf{x}_k,t) \right]f_{i,j}(\mathbf{x}_j,\mathbf{x}_k,t).
\label{eq:fij-relation-III}
\end{eqnarray}
While the first condition arises from requiring the plaquettes in the $i-j$ plane to have their desired value when $x_i=L-a_s$, the last two conditions arise from setting the value of plaquettes in the $i-k$ and $i-0$ plane to their desired values.
Furthermore, by studying carefully the value of the plaquettes in the $i-j$ and $i-k$ planes when $\mathbf{x}_i=L-a_s$, and given the modified links in Eq. (\ref{eq:Ui-modified-general-compl}), one arrives at the following conditions on $f_{i,0}$
\begin{eqnarray}
&& \left[ A_0(\mathbf{x}_i=0,\mathbf{x}_j+a_s,\mathbf{x}_k,t)-\widetilde{A}_0(\mathbf{x}_i=L,\mathbf{x}_j+a_s,\mathbf{x}_k,t) \right]f_{i,0}(\mathbf{x}_j+a_s,\mathbf{x}_k,t)
\nonumber\\
&& \qquad \qquad \qquad \qquad \qquad \qquad =\left[ A_0(\mathbf{x}_i=0,\mathbf{x}_j,\mathbf{x}_k,t)-\widetilde{A}_0(\mathbf{x}_i=L,\mathbf{x}_j,\mathbf{x}_k,t) \right]f_{i,0}(\mathbf{x}_j,\mathbf{x}_k,t),
\label{eq:fi0-relation-II}
\end{eqnarray}
\begin{eqnarray}
&& \left[ A_0(\mathbf{x}_i=0,\mathbf{x}_j,\mathbf{x}_k+a_s,t)-\widetilde{A}_0(\mathbf{x}_i=L,\mathbf{x}_j,\mathbf{x}_k+a_s,t) \right]f_{i,0}(\mathbf{x}_j,\mathbf{x}_k+a_s,t)
\nonumber\\
&& \qquad \qquad \qquad \qquad \qquad \qquad=\left[ A_0(\mathbf{x}_i=0,\mathbf{x}_j,\mathbf{x}_k,t)-\widetilde{A}_0(\mathbf{x}_i=L,\mathbf{x}_j,\mathbf{x}_k,t) \right]f_{i,0}(\mathbf{x}_j,\mathbf{x}_k,t).
\label{eq:fi0-relation-III}
\end{eqnarray}
These conditions add to the condition in Eq. (\ref{eq:fi0-relation-I}) and must be satisfied simultaneously to obtain a consistent solution for $f_{i,0}$. Similarly, if the value of the plaquette in the $0-j$ and $0-k$ planes are considered when $t=T-a_t$, one arrives at
\begin{eqnarray}
&& \left[A_i(\mathbf{x}_i,\mathbf{x}_j+a_s,\mathbf{x}_k,t=0)-\widetilde{A}_i(\mathbf{x}_i,\mathbf{x}_j+a_s,\mathbf{x}_k,t=T)\right]f_{0,i}(\mathbf{x}_i,\mathbf{x}_j+a_s,\mathbf{x}_k)
\nonumber\\
&& \qquad \qquad \qquad \qquad \qquad \qquad \qquad \qquad \qquad \qquad =\left[ A_i(\mathbf{x},t=0)-\widetilde{A}_i(\mathbf{x},t=T) \right] f_{0,i}(\mathbf{x}),
\label{eq:f0i-relation-II}
\end{eqnarray}
\begin{eqnarray}
&& \left[A_i(\mathbf{x}_i,\mathbf{x}_j,\mathbf{x}_k+a_s,t=0)-\widetilde{A}_i(\mathbf{x}_i,\mathbf{x}_j,\mathbf{x}_k+a_s,t=T)\right]f_{0,i}(\mathbf{x}_i,\mathbf{x}_j,\mathbf{x}_k)
\nonumber\\
&& \qquad \qquad \qquad \qquad \qquad \qquad \qquad \qquad \qquad \qquad =\left[ A_i(\mathbf{x},t=0)-\widetilde{A}_i(\mathbf{x},t=T) \right] f_{0,i}(\mathbf{x}),
\label{eq:f0i-relation-III}
\end{eqnarray}
which supplement the previous condition on $f_{0,i}$ in Eq. (\ref{eq:f0i-relation-I}). 

With these modified links, one can see that, in addition to the three QCs in Eq. (\ref{eq:QC-0i-general}) for $i=1,2,3$, three more QCs arise
\begin{eqnarray}
\left[\prod_{\mathbf{x}_i=0}^{L-a_s} e^{-ie\hat{Q}\left[ A_i(\mathbf{x}_i,\mathbf{x}_j=0,\mathbf{x}_k,t)-\widetilde{A}_i(\mathbf{x}_i,\mathbf{x}_j=L,\mathbf{x}_k,t) \right]a_s}\right]
\left[\prod_{\mathbf{x}_j=0}^{L-a_s} e^{ie\hat{Q}\left[ A_j(\mathbf{x}_i=0,\mathbf{x}_j,\mathbf{x}_k,t)-\widetilde{A}_j(\mathbf{x}_i=L,\mathbf{x}_j,\mathbf{x}_k,t) \right]a_s}\right]=1,
\label{eq:QC-ij-general}
\end{eqnarray}
by requiring that the desired value of the plaquette located at $\mathbf{x}_i=L-a_s$ and $\mathbf{x}_j=L-a_s$ is generated. Eq. (\ref{eq:QC-ij-general}) is the statement that the net flux of the magnetic field through the $i-j$ plane must be quantized, $\Phi_{(i,j)}^{(\mathbf{B}),\text{net}}(\mathbf{x}_k,t)=\frac{2\pi n}{e\hat{Q}}$ with $n \in \mathbb{Z}$. Therefore, if the flux is dependent on the $\mathbf{x}_k$ ($k \neq i,j$) and $t$ coordinates, this condition can not be satisfied in general.

Equations above for the $f_{\mu,\nu}$ functions must be satisfied simultaneously to ensure that the proper values of the elementary plaquettes are produced throughout the lattice. However, it is not guaranteed that, for any given $A$ field, these equations possess consistent solutions for each $f_{\mu,\nu}$. To clarify this point, consider the $f_{i,0}$ function which must be obtained recursively from Eqs. (\ref{eq:fi0-relation-I}), (\ref{eq:fi0-relation-II}) and (\ref{eq:fi0-relation-III}). It is straightforward to see that these equations are consistent with each other only when $A_0$ depends solely on the $\mathbf{x}_i$ and $t$ coordinates.  In general, there exists a valid $f_{\mu,\nu}$ only if $A_{\nu}$ solely depends on $x_{\mu}$ and $x_{\nu}$ coordinates. Note that if $A_{\nu}$ is independent of the $x_{\mu}$ coordinate, no discontinuity occurs in the value of plaquette in the $\mu-\nu$ plane when $x_{\mu}=L_{\mu}-a_{\mu}$ ($L_{\mu}=T$ and $a_{\mu}=a_t$ for $\mu=0$ while $L_{\mu}=L$ and $a_{\mu}=a_s$ for $\mu=i$). As a result no $f_{\mu,\nu}$ needs to be introduced to guarantee periodicity.\footnote{Another way to ensure that the desired values of the plaquettes are produced adjacent to the boundaries of the lattice is to enforce a set of \emph{micro QCs}. These QCs can be deduced by setting the extra factor in the value of plaquettes near the boundaries equal to one (without requiring any gauge link to be modified). For example, one can set the coordinate-dependent factors $e^{ie\hat{Q}\left[\widetilde{A}_0(L,\mathbf{x}_j,\mathbf{x}_k,t)-A_0(0,\mathbf{x}_j,\mathbf{x}_k,t)\right]a_t}$ in Eq. (\ref{eq:plq-0i-Lminus1}) and  $e^{-ie\hat{Q}\left[\widetilde{A}_i(\mathbf{x}_i,\mathbf{x}_j,\mathbf{x}_k,T)-A_i(\mathbf{x}_i,\mathbf{x}_j,\mathbf{x}_k,0)\right]a_s}$ in Eq. (\ref{eq:plq-0i-Tminus1}) equal to 1, such that not only the correct values of plaquettes in the $0-i$ plane at  $0 \leq \mathbf{x}_i < L-a_s,~t=T-a_t$ and $\mathbf{x}_i=L-a_s,~0 \leq t < T-a_t$ are produced, but also the correct value of plaquette at the far corner of the lattice, i.e., at $\mathbf{x}_i=L-a_s,~t=T-a_t$, is produced, without the emergence of an extra factor as in Eq. (\ref{eq:plq-0i-Lminus1-Tminus1}). Clearly, the extra factor in Eq. (\ref{eq:plq-0i-Lminus1-Tminus1}), which is a product of all the coordinate-dependent phase factors above, is automatically equal to $1$ because of the micro QCs. Unfortunately, this procedure will not alway work due to the coordinate-dependence of the new conditions (note that even if the gauge fields are chosen to be independent of the transverse coordinate, the QCs still carry a longitudinal coordinate dependence). On the lattice, where space and time coordinates are discretized, gauge fields with simple rational dependences on the coordinates could allow such micro QCs to be satisfied. However, such conditions are more restrictive on the background field parameters than the QC derived in this section on the total flux of the field, leading to large quanta of background fields that is not desired for most applications.} Interestingly, all such conditions on the space-time dependence of $A_{\mu}$ can be shown to be consistent with the statement that the net electric or magnetic flux through any plane on the four-dimensional lattice (a closed surface in the toroidal geometry) must be space-time independent. This is exactly the condition we deduced by examining the QCs in Eq. (\ref{eq:QC-0i-general}) and (\ref{eq:QC-ij-general}). In the next section, we present the setup for the implementation of several chosen background fields on a periodic lattice and and will specify the corresponding QCs.

\section{EXAMPLES: PERIODIC IMPLEMENTATION OF SELECTED BACKGROUND FIELDS ON A HYPERCUBIC LATTICE DEDUCED FROM THE GENERAL CASE
\label{sec:Examples} 
}
\noindent
The examples that follow provide a setup for the implementation of selected background electric and/or magnetic fields that preserves the periodicity of the lattice calculation. These are deduced from the general case of the previous section, the results of which will be summarized here for convenience. In order for the background $U(1)$ gauge links to be implemented periodically, they must be introduced as
\begin{eqnarray}
U^{(\text{QCD})}_{\mu}(x) & \rightarrow & U_{\mu}^{(\text{QCD})}(x)  \times e^{ie\hat{Q}A_{\mu}(x)a_{\mu}} \times \prod_{\nu \neq \mu} e^{ie\hat{Q}\left[A_{\nu}(x_{\mu}=0,x_{\nu})-\widetilde{A}_{\nu}(x_{\mu}=L_{\mu},x_{\nu})\right]f_{\mu,\nu}(x_{\nu}) \times \delta_{x_{\mu},L_{\mu}-a_{\mu}}},
\label{eq:U-mu-general}
\end{eqnarray}
where $\mu$ and $\nu$ assume distinct values. If each $A_{\mu}$ depends only on $x_{\mu}$ and $x_{\nu}$ coordinates, there exist functions $f_{\mu,\nu}$ that satisfy the following recursive relation on the lattice, 
\begin{eqnarray}
&& \left[ A_{\nu}(x_{\mu}=0,x_{\nu}+a_{\nu})-\widetilde{A}_{\nu}(x_{\mu}=L_{\mu},x_{\nu}+a_{\nu}) \right]f_{\mu,\nu}(x_{\nu}+a_{\nu})=
\nonumber\\
&& ~~~~~~ \qquad \qquad \qquad \qquad \qquad \qquad \left[ A_{\nu}(x_{\mu}=0,x_{\nu})-\widetilde{A}_{\nu}(x_{\mu}=L_{\mu},x_{\nu}) \right]\left(f_{\mu,\nu}(x_{\nu})+a_{\nu}\right).
\label{eq:f-mu-nu-relation}
\end{eqnarray}
$f_{\mu,\nu}$ is vanishing if $A_{\nu}$ is independent of $x_{\mu}$.\footnote{In cases where $A_{\nu}$ does not depend on the $x_{\nu}$ coordinate, as is the case in most of the examples in this section, the $f_{\mu,\nu}$ satisfies the relation $f_{\mu,\nu}(x_{\nu}+a_{\nu})=f_{\mu,\nu}(x_{\nu})+a_{\nu}$ with the solution $f_{\mu,\nu}(x_{\nu})=x_{\nu}$, once one sets $f_{\mu,\nu}(0)=0$. If $A_{\nu}$ depends on both $x_{\mu}$ and $x_{\nu}$ coordinates, it is possible to transform to a gauge where $A_{\nu}$ does not depend on $x_{\nu}$, as long as the condition $\partial_{\mu}\partial_{\nu}F^{\mu \nu}=0$ is satisfied, where $F^{\mu \nu}$ is the EM field strength tensor and no summation over $\mu$ and $\nu$ is assumed.} Under the conditions specified, the net electric or magnetic flux through any plane in the continuum limit is constant and must be quantized. On the lattice these QCs read
\begin{eqnarray}
\left[\prod_{x_{\mu}=0}^{L_{\mu}-a_{\mu}} e^{-ie\hat{Q}\left[ A_{\mu}(x_{\mu},x_{\nu}=0)-\widetilde{A}_{\mu}(x_{\mu},x_{\nu}=L_{\nu}) \right]a_{\mu}}\right]
\left[\prod_{x_{\nu}=0}^{L_{\nu}-a_{\nu}} e^{ie\hat{Q}\left[ A_{\nu}(x_{\mu}=0,x_{\nu})-\widetilde{A}_{\nu}(x_{\mu}=L_{\mu},x_{\nu}) \right]a_{\nu}}\right]=1.
\label{eq:QC-mu-nu-general}
\end{eqnarray}
All the space-time dependences in the following examples must be understood through the floor-function prescription of Sec. \ref{sec:Special-case}.

\subsection*{Example I: A constant electric field in the $\mathbf{x}_3$ direction
\label{subsec:E1-II} 
}
We have already discussed this case in Sec. \ref{sec:Special-case}. Here we choose a different gauge than that taken in Sec. \ref{sec:Special-case},
\begin{eqnarray}
A_{\mu}=(A_0,-\mathbf{A})=\left(0,0,0,a_1t\right),
\end{eqnarray}
which produces an electric field in the $\mathbf{x}_3$ direction, $\mathbf{E}=a_1 \hat{\mathbf{x}}_3$. The nontrivial gauge links that produce this background field are
\begin{eqnarray}
U^{(\text{QCD})}_0 & \rightarrow & U_0^{(\text{QCD})} \times 
e^{-ie\hat{Q} a_1 T f_{0,3}(\mathbf{x}_3) \times \delta_{t,T-a_t}},
\\
U^{(\text{QCD})}_3 & \rightarrow & U_3^{(\text{QCD})} \times e^{ie\hat{Q}a_1ta_s},
\end{eqnarray}
where $f_{0,3}$ satisfies $f_{0,3}(\mathbf{x}_3+a_s)=f_{0,3}(\mathbf{x}_3)+a_s$, with the solution $f_{0,3}(\mathbf{x}_3)=\mathbf{x}_3$ once one sets $f_{0,3}(0)=0$.\footnote{Note that the initial value of the function is arbitrary as it drops out of the value of plaquettes.} There is only one QC in this case,
\begin{eqnarray}
\prod_{\mathbf{x}_3=0}^{L-a_s} e^{-ie\hat{Q}a_1Ta_s}=\left[ e^{-ie\hat{Q}a_1T a_s} \right]^{\frac{L}{a_s}}=e^{-ie\hat{Q}a_1TL}=1,
\label{eq:case-1-QC-I}
\end{eqnarray}
which constrains the value of the field strength, $a_1$,
\begin{eqnarray}
a_1= \frac{2\pi n}{e\hat{Q}TL},
\end{eqnarray}
with $n \in \mathbb{Z}$.

\subsection*{Example II: A constant magnetic field in the $\mathbf{x}_3$ direction
\label{subsec:E2-II} 
}
One can pick the following gauge potential
\begin{eqnarray}
A_{\mu}=(A_0,-\mathbf{A})=\left(0,\frac{b_1}{2} \mathbf{x}_2,-\frac{b_1}{2} \mathbf{x}_1,0\right),
\end{eqnarray}
to generate a uniform magnetic field in the $\mathbf{x}_3$ direction, $\mathbf{B}=b_1 \hat{\mathbf{x}}_3$. The nontrivial gauge links that are required to implement this background field are
\begin{eqnarray}
U^{(\text{QCD})}_1 & \rightarrow & U_1^{(\text{QCD})} \times e^{\frac{i}{2}e\hat{Q}b_1\mathbf{x}_2 a_s} \times
e^{\frac{i}{2}e\hat{Q} b_1 L f_{1,2}(\mathbf{x}_2) \times \delta_{\mathbf{x}_1,L-a_s}},
\\
U^{(\text{QCD})}_2 & \rightarrow & U_2^{(\text{QCD})} \times e^{-\frac{i}{2}e\hat{Q}b_1\mathbf{x}_1 a_s} \times
e^{-\frac{i}{2}e\hat{Q} b_1 L f_{2,1}(\mathbf{x}_1) \times \delta_{\mathbf{x}_2,L-a_s}}.
\end{eqnarray}
The conditions on $f_{1,2}$ and $f_{2,1}$ are respectively, $f_{1,2}(\mathbf{x}_2+a_s)=f_{1,2}(\mathbf{x}_2)+a_s$ and $f_{2,1}(\mathbf{x}_1+a_s)=f_{2,1}(\mathbf{x}_1)+a_s$, with the solutions $f_{1,2}=\mathbf{x}_2$ and $f_{2,1}=\mathbf{x}_1$, once the initial values of the functions are set to zero. There is one QC for this choice of the field
\begin{eqnarray}
\left[\prod_{\mathbf{x}_1=0}^{L-a_s} e^{\frac{i}{2}e\hat{Q}b_1 L a_s}\right]
\left[\prod_{\mathbf{x}_2=0}^{L-a_s} e^{\frac{i}{2}e\hat{Q}b_1 L a_s}\right]=\left[ e^{ie\hat{Q}b_1 L a_s} \right]^{\frac{L}{a_s}}=e^{ie\hat{Q}b_1L^2}=1,
\label{eq:case-2-QC-I}
\end{eqnarray}
which constrains the strength of the magnetic field parameter,
\begin{eqnarray}
b_1= \frac{2\pi n}{e\hat{Q}L^2},
\end{eqnarray}
with $n \in \mathbb{Z}$.

\subsection*{Example III: A space-dependent magnetic field and a constant electric field
\label{subsec:E3-II} 
}
When interested in extracting the spin polarizabilities of nucleons, one can choose the following gauge potential
\begin{eqnarray}
A_{\mu}=(A_0,-\mathbf{A})=\left(0,b_2 \mathbf{x}_1\mathbf{x}_2,0,a_2t\right),
\end{eqnarray}
to produce a space-dependent magnetic field, $\mathbf{B}=b_2 \mathbf{x}_1 \hat{\mathbf{x}}_3$, as well as a constant electric field, $\mathbf{E}=a_2 \hat{\mathbf{x}}_3$. These background fields generate a nonvanishing interaction proportional to $\frac{1}{2}\sigma_i (\nabla_i\mathbf{B}_j+\nabla_j\mathbf{B}_i)\mathbf{E}_j$ (with $\sigma_i$ denoting Pauli matrices) in the nonrelativistic Hamiltonian of the spin-$\frac{1}{2}$ hadron in external fields, which gives access to the $\gamma_{E_1M_2}$ spin polarizability of the hadron, see Ref. \cite{Detmold:2006vu}. To produce this electric field on a periodic lattice, the nontrivial gauge links to be implemented are
\begin{eqnarray}
U^{(\text{QCD})}_0 & \rightarrow & U_0^{(\text{QCD})} \times
e^{-ie\hat{Q} a_2 T f_{0,3}(\mathbf{x}_3) \times \delta_{t,T-a_t}},
\\
U^{(\text{QCD})}_1 & \rightarrow & U_1^{(\text{QCD})} \times e^{ie\hat{Q}b_2\mathbf{x}_1\mathbf{x}_2a_s},
\\
U^{(\text{QCD})}_2 & \rightarrow & U_2^{(\text{QCD})} \times e^{-ie\hat{Q}b_2 \mathbf{x}_1 L f_{2,1}(\mathbf{x}_1) \times \delta_{\mathbf{x}_2,L-a_s}},
\\
U^{(\text{QCD})}_3 & \rightarrow & U_3^{(\text{QCD})} \times e^{ie\hat{Q}a_2ta_s}.
\end{eqnarray}
The $f_{0,3}$ and $f_{2,1}$ functions satisfy recursive relations $f_{0,3}(\mathbf{x}_3+a_s)=f_{0,3}(\mathbf{x}_3)+a_s$ and $f_{2,1}(\mathbf{x}_1+a_s)=\frac{\mathbf{x}_1}{\mathbf{x}_1+a_s}\left(f_{2,1}(\mathbf{x}_1)+a_s\right)$, respectively, with solutions $f_{0,3}(\mathbf{x}_3)=\mathbf{x}_3$ and $f_{2,1}(\mathbf{x}_1)=\frac{\mathbf{x}_1-a_s}{2}$ for $\mathbf{x}_1>0$, once the initial values of the functions are set to zero. The only QCs are
\begin{eqnarray}
\prod_{\mathbf{x}_1=0}^{L-a_s} e^{ie\hat{Q}b_2 \mathbf{x}_1 La_s}=
e^{\frac{i}{2}e\hat{Q}b_2 L^2(L-a_s)}=e^{\frac{i}{2}e\hat{Q}b_2 L^3(1-\frac{a_s}{L})}=1,
\label{eq:case-5-QC-I}
\end{eqnarray}
\begin{eqnarray}
\prod_{\mathbf{x}_3=0}^{L-a_s} e^{-ie\hat{Q}a_2Ta_s}=\left[ e^{-ie\hat{Q}a_2Ta_s} \right]^{\frac{L}{a_s}}=e^{-ie\hat{Q}a_2TL}=1,
\label{eq:case-5-QC-II}
\end{eqnarray}
and therefore
\begin{eqnarray}
b_2= \frac{4\pi n}{e\hat{Q} L^3(1-\frac{a_s}{L})},~~~a_2= \frac{2\pi n'}{e\hat{Q}TL},
\end{eqnarray}
with $n,n' \in \mathbb{Z}$. Note that if we had only quantized the flux of the magnetic field in the continuum limit, we would have introduced a deviation from periodicity on the lattice of $\mathcal{O}(\frac{a_s}{L})$. To avoid such discretization errors one must quantize the fields according to Eq. (\ref{eq:QC-mu-nu-general}), thereby respecting the lattice geometry.

\subsection*{Example IV: A time-dependent electric field
\label{subsec:E4-II} 
}
Another spin polarizability of nucleon can be accessed via the background gauge potential
\begin{eqnarray}
A_{\mu}=(A_0,-\mathbf{A})=\left(0,\frac{1}{2}a_3 t^2,a_4t,0\right),
\end{eqnarray}
which produces a time-dependent electric field, $\mathbf{E}=a_3t \hat{\mathbf{x}}_1+a_4 \hat{\mathbf{x}}_2$.\footnote{Any time-dependent magnetic field will necessarily generate a time-dependent flux through the transverse plane. This means that one can not implement a periodic background field with the method of this work to access the $\gamma_{M_1M_1}$ spin polarizability of hadron. This is, up to a numerical factor, the coefficient of the $\bm{\sigma} \cdot \mathbf{B} \times \dot{\mathbf{B}}$ term in the nonrelativistic Hamiltonian of the spin-$\frac{1}{2}$ hadron in an external magnetic field.} This background field generates a nonvanishing interaction proportional to $\bm{\sigma} \cdot \mathbf{E} \times \dot{\mathbf{E}}$ in the nonrelativistic Hamiltonian of the spin-$\frac{1}{2}$ hadron in external fields, which gives access to the $\gamma_{E_1E_1}$ spin polarizability of the hadron, see Ref. \cite{Detmold:2006vu}. To produce this electric field on a periodic lattice, the nontrivial gauge links to be implemented are
\begin{eqnarray}
U^{(\text{QCD})}_0 & \rightarrow & U_0^{(\text{QCD})} \times
e^{-\frac{i}{2}e\hat{Q} a_3 T^2 f_{0,1}(\mathbf{x}_1) \times \delta_{t,T-a_t}} \times 
e^{-ie\hat{Q} a_4 T f_{0,2}(\mathbf{x}_2) \times \delta_{t,T-a_t}}, 
\\
U^{(\text{QCD})}_1 & \rightarrow & U_1^{(\text{QCD})} \times e^{\frac{i}{2}e\hat{Q}a_3t^2a_s}, 
\\
U^{(\text{QCD})}_2 & \rightarrow & U_2^{(\text{QCD})} \times e^{ie\hat{Q} a_4 t a_s}.
\end{eqnarray}
The $f_{0,1}$ and $f_{0,2}$ functions satisfy recursive relations $f_{0,1}(\mathbf{x}_1+a_s)=f_{0,1}(\mathbf{x}_1)+a_s$ and $f_{0,2}(\mathbf{x}_2+a_s)=f_{0,2}(\mathbf{x}_2)+a_s$, respectively, with trivial solutions $f_{0,1}(\mathbf{x}_1)=\mathbf{x}_1$ and $f_{0,2}(\mathbf{x}_2)=\mathbf{x}_2$, once the initial values of the functions are set to zero. The only QCs are
\begin{eqnarray}
\prod_{\mathbf{x}_1=0}^{L-a_s} e^{\frac{i}{2}e\hat{Q}a_3T^2a_s}=\left[ e^{\frac{i}{2}e\hat{Q}a_3T^2a_s} \right]^{\frac{L}{a_s}}=e^{\frac{i}{2}e\hat{Q}a_3T^2L}=1,
\label{eq:case-6-QC-I}
\end{eqnarray}
\begin{eqnarray}
\prod_{\mathbf{x}_2=0}^{L-a_s} e^{ie\hat{Q}a_4Ta_s}=\left[ e^{ie\hat{Q}a_4Ta_s} \right]^{\frac{L}{a_s}}=e^{ie\hat{Q}a_4TL}=1,
\label{eq:case-6-QC-III}
\end{eqnarray}
and therefore
\begin{eqnarray}
a_3= \frac{4\pi n}{e\hat{Q}T^2L},~~~a_4= \frac{2\pi n'}{e\hat{Q}TL},
\end{eqnarray}
with $n,n' \in \mathbb{Z}$.

\subsection*{Example V: A space-dependent electric field and a constant magnetic field
\label{subsec:E5-II} 
}
A third spin polarizability of nucleon can be accessed with the following gauge potential
\begin{eqnarray}
A_{\mu}=(A_0,-\mathbf{A})=\left(-\frac{a_5}{2}\mathbf{x}_2^2,0,0,b_3\mathbf{x}_1\right),
\end{eqnarray}
which produces a space-dependent electric field, $\mathbf{E}=a_5\mathbf{x}_2\hat{\mathbf{x}}_2$ and a constant magnetic field, $\mathbf{B}=b_3\hat{\mathbf{x}}_2$. This background field generates a nonvanishing interaction proportional to $\frac{1}{2}\sigma_i (\nabla_i\mathbf{E}_j+\nabla_j\mathbf{E}_i)\mathbf{B}_j$ in the nonrelativistic Hamiltonian of the spin-$\frac{1}{2}$ hadron in external fields, and so gives access to the $\gamma_{M_1E_2}$ spin polarizability of the hadron, see Ref. \cite{Detmold:2006vu}. To produce this electric field on a periodic lattice, the nontrivial gauge links to be implemented are
\begin{eqnarray}
U^{(\text{QCD})}_0 & \rightarrow & U_0^{(\text{QCD})} \times
e^{-\frac{i}{2}e\hat{Q} a_5 \mathbf{x}_2^2 a_t}, 
\\
U^{(\text{QCD})}_1 & \rightarrow & U_1^{(\text{QCD})} \times e^{-ie\hat{Q}b_3Lf_{1,3}(\mathbf{x}_3) \times \delta_{\mathbf{x}_1,L-a_s}}, 
\\
U^{(\text{QCD})}_2 & \rightarrow & U_2^{(\text{QCD})} \times  e^{\frac{i}{2}e\hat{Q}a_5L^2f_{2,0}(t) \times \delta_{\mathbf{x}_2,L-a_s}}, 
\\
U^{(\text{QCD})}_3 & \rightarrow & U_3^{(\text{QCD})} \times e^{ie\hat{Q} b_3 \mathbf{x}_1 a_s}.
\end{eqnarray}
The $f_{1,3}$ and $f_{2,0}$ functions satisfy recursive relations $f_{1,3}(\mathbf{x}_3+a_s)=f_{1,3}(\mathbf{x}_3)+a_s$ and $f_{2,0}(t+a_t)=f_{2,0}(t)+a_t$, respectively, with trivial solutions $f_{1,3}(\mathbf{x}_3)=\mathbf{x}_3$ and $f_{2,0}(t)=t$, once the initial values of the functions are set to zero. The only QCs are
\begin{eqnarray}
\prod_{t=0}^{T-a_t} e^{-\frac{i}{2}e\hat{Q}a_5L^2a_t}=\left[ e^{-\frac{i}{2}e\hat{Q}a_5L^2a_t} \right]^{\frac{T}{a_t}}=e^{-\frac{i}{2}e\hat{Q}a_5L^2T}=1,
\label{eq:case-6-QC-I}
\end{eqnarray}
\begin{eqnarray}
\prod_{\mathbf{x}_3=0}^{L-a_s} e^{-ie\hat{Q}b_3La_s}=\left[ e^{-ie\hat{Q}b_3La_s} \right]^{\frac{L}{a_s}}=e^{-ie\hat{Q}b_3L^2}=1,
\label{eq:case-6-QC-III}
\end{eqnarray}
and therefore
\begin{eqnarray}
a_5= \frac{4\pi n}{e\hat{Q}L^2T},~~~b_3= \frac{2\pi n'}{e\hat{Q}L^2},
\end{eqnarray}
with $n,n' \in \mathbb{Z}$.

\subsection*{Example VI: A plane-wave electric field
\label{subsec:E6-II} 
}
As suggested in Ref. \cite{Detmold:2004kw}, background EM plane waves can be used to calculate the off-forward matrix elements of current operators between hadronic states, enabling an extraction of hadronic form factors.\footnote{In general, any space-time inhomogeneity in the background field gives access to such off-forward quantities due to injecting energy and/or momentum. Plane-wave background fields have the advantage of isolating contributions with a given momentum transfer.} Additionally, a recent proposal in Ref. \cite{Bali:2015msa} demonstrates the advantage of a plane-wave background field in evaluating the hadronic vacuum polarization function with lattice QCD. This approach proceeds by constraining the polarization function using the susceptibilities with respect to the background magnetic field amplitude at specific momenta. Due to the condition on the space-time dependence of the flux of the background field in each plane, our periodic implementation of EM plane waves will be limited to fields with certain Fourier modes. For example, an electric field of the form $\mathbf{E}=e^{i \mathbf{q} \cdot \mathbf{x}} \hat{\mathbf{x}}_3$ with $\mathbf{q}_i \neq 0$ for all $i=1,2,3$ will generate a coordinate-dependent flux. We can however generate an electric field of the form, e.g., $\mathbf{E}=a_6e^{i q_3 \mathbf{x}_3} \hat{\mathbf{x}}_3$, from the following gauge potential  
\begin{eqnarray}
A_{\mu}=(A_0,-\mathbf{A})=\left(\frac{ia_6}{q_3}e^{i q_3 \mathbf{x}_3},0,0,0\right),
\label{eq:A-palne-wave}
\end{eqnarray}
with a constant flux through the $0-3$ plane. Since the form factors in the continuum (infinite-volume) limit are rotationally invariant, the constraints on the components of the momentum transfer vector in this setup will not prevent one from accessing the form factors at any $q^2$ values. The only limitation on the (magnitude) of the transferred momenta may arise from the implementation of fields on a periodic lattice as will be deduced below.

To produce the electric field chosen above on a periodic lattice, the nontrivial gauge links to be implemented are
\begin{eqnarray}
U^{(\text{QCD})}_0 & \rightarrow & U_0^{(\text{QCD})} \times
e^{-\frac{e\hat{Q}a_6}{q_3}e^{i q_3 \mathbf{x}_3} a_t},
\\
U^{(\text{QCD})}_3 & \rightarrow & U_3^{(\text{QCD})} \times e^{-\frac{e\hat{Q}a_6}{q_3}(1-e^{i q_3L})f_{3,0}(t) \times \delta_{\mathbf{x}_3,L-a_s}},
\label{eq:U1-plane-wave}
\end{eqnarray}
where $f_{3,0}$ satisfies the recursive relation $f_{3,0}(t+a_t)=f_{3,0}(t)+a_t$ with trivial solution $f_{3,0}(t)=t$, once the initial value of the function is set to zero. The only QC is
\begin{eqnarray}
\prod_{t=0}^{T-a_t} e^{-\frac{e\hat{Q}a_6}{q_3}(1-e^{i q_3L})a_t}=\left[ e^{-\frac{e\hat{Q}a_6}{q_3}(1-e^{i q_3L})a_t} \right]^{\frac{T}{a_t}}=e^{-\frac{e\hat{Q}a_6}{q_3}(1-e^{i q_3L})T}=1.
\label{eq:case-6-QC-I}
\end{eqnarray}
For any arbitrary value of the field amplitude parameter, $a_6$, this QC can be satisfied with
\begin{eqnarray}
q_3=\frac{2\pi n}{L},
\label{eq:q3-QC}
\end{eqnarray}
with $n \in \mathbb{Z}$. This constraint on $q_3$ means that the additional $U(1)$ phase factor in Eq. (\ref{eq:U1-plane-wave}) is equal to unity. It also means that with a background field method, the EM form factors can only be accessed at quantized values of momentum transfer, the situation which is also encountered when form factors are calculated through a direct evaluation of hadronic matrix elements on the lattice. However, one could allow for nonqunatized $q_3$ values by placing conditions on the real part, $a_6^{(r)}$, and imaginary part, $a_6^{(i)}$, of $a_6$. Indeed, by requiring
\begin{eqnarray}
a_6^{(i)}=\frac{\pi q_3 n'}{e\hat{Q}T},~a_6^{(r)}=-\frac{\sin (q_3 L)}{1-\cos(q_3L)}a_6^{(i)},
\end{eqnarray}
with $n' \in \mathbb{Z}$ and $q_3 \neq \frac{2\pi n}{L}$ for $n \in \mathbb{Z}$, the QC in Eq. (\ref{eq:case-6-QC-I}) is satisfied. As we already saw, for $q_3 = \frac{2\pi n}{L}$ the QC in Eq. (\ref{eq:case-6-QC-I}) trivially holds.\footnote{A plane-wave background gauge field as in Eq. (\ref{eq:A-palne-wave}) has a nonvanishing imaginary piece. It therefore results in a non-Hermitian fermionic determinant (after integrating out the quark fields) which can hinder the probabilistic evaluation of the associated path integral (analogous to the effect of a nonvanishing chemical potential). For isovector quantities, one can avoid this \emph{sign} problem imposed  by such plane-wave gauge field by only implementing the background fields on the valence quarks. For isoscalar quantities, where there are disconnected contributions to the matrix elements of the EM current, contributions from a charged quark sea can not be ignored. For these quantities, where a full imposition of the background fields on the valence and sea quarks is required, the computational cost associated with a non-Hermitian measure in the Monte Carlo sampling of the path integral can be controlled by tuning the amplitude of the gauge field to be small. Note that the amplitude of the field can be made arbitrarily small (but nonvanishing) only if $q_3$ is quantized according to Eq. (\ref{eq:q3-QC}). Alternatively, one can implement only real oscillatory functions, i.e. $\sin(q_3 \mathbf{x}_3)$ or $\cos(q_3 \mathbf{x}_3)$, to access the desired off-forward matrix elements.}
 The result of this latter case is indeed consistent with the periodic implementation of an oscillatory magnetic field (through $\sin$ and $\cos$ functions) in Ref. \cite{Bali:2015msa} in which the Fourier modes of the applied field are taken to be quantized.

\section{Conclusion
\label{sec:Conclusion} 
}
\noindent
The introduction of classical electromagnetic fields in lattice quantum chromodynamics calculations provides a powerful technique to obtain a variety of electromagnetic properties of hadrons and nuclei. To extend the utility of this technique, the current implementations of uniform background fields \cite{Bernard:1982yu, Martinelli:1982cb, Fiebig:1988en, Christensen:2004ca, Lee:2005ds, Lee:2005dq, Detmold:2006vu, Aubin:2008qp, Detmold:2009dx, Detmold:2010ts, Primer:2013pva, Lujan:2014kia, Beane:2014ora, Beane:2015yha} have been extended in this paper to the case of background fields that have temporal and/or spatial nonuniformities. Such field configurations provide access to static and quasi-static properties of hadrons and nuclei, such as their higher EM moments and polarizabilities as well as their charge radii. They also provide means to directly extract EM from factors as energy and momentum are injected into the hadronic system immersed in appropriate external fields \cite{Detmold:2004kw}. Such possibilities can be explored in upcoming lattice QCD calculations once the corresponding background fields required for each quantity is consistently implemented on the particular lattice geometry. Periodic boundary conditions are widely used in lattice QCD calculations, and given that such boundary conditions typically result in a simpler hadronic theory in general to be matched to lattice QCD calculations, it is important to perform the background field calculation with the imposition of these boundary conditions. In this work, we have considered the most general space-time dependent $U(1)$ gauge fields imposed on QCD gauge configurations, and have shown that under certain conditions on the space-time dependence of the fields, a periodic implementation of background $U(1)$ gauge fields is possible.

To make the discussions more transparent before getting into formalities of the general case, we have first presented the special case of an electric field generated through a scalar potential with an arbitrary dependence on one spatial coordinate. The necessary modifications to the naive $U(1)$ links adjacent to the boundary of the lattice are obtained by ensuring the expected values of the elementary plaquettes are produced throughout the lattice. These expected values are nothing but the $U(1)$ phases corresponding the flux of the electric field in the continuum limit through surface areas $a_t \times a_s$ at each space-time point on the lattice, where $a_t$ and $a_s$ refer to lattice spacings in temporal and spatial directions, respectively. Additionally, a quantization condition is obtained that ensures that the flux of the electric field is quantized. From this special case, two examples of a uniform, and a linearly varying electric field in space, are constructed. We have numerically confirmed that only the periodic  implementation of gauge fields, according to the prescription proposed, gives rise to smooth correlation functions for neutral pions across the boundary of the lattice. 

For the general case of gauge fields with arbitrary space-time dependences, one can follow the same procedure as that of the special case above, except that obtaining the modified links near the boundary is more involved. This is simply because of  the fact that when a component of the gauge field depends on more than one space-time coordinate, fixing the values of plaquettes in one plane to their desired values (by modifying the links adjacent to the boundary) can potentially affect the values of the plaquettes in other planes. By carefully accounting for such possibilities, we have derived a set of equations for the functions that need to be introduced in the modified links and have discussed the conditions on classical fields that guarantee the existence of solutions for these equations. We have further shown that these conditions are equivalent to the statement that the flux of the electric and magnetic fields through each plane of the lattice must be coordinate independent and quantized. The latter is a condition that must be met to ensure the expected value of the plaquette at the far corner of the lattice in each plane is produced. In a parallel approach, we have shown with details in the appendix that these conditions arise from a more general class of boundary conditions, namely \emph{electro/magneto-periodic boundary conditions} \cite{'tHooft1979141, 'tHooft1981, vanBaal1982, Smit:1986fn, Damgaard:1988hh, Rubinstein:1995hc, AlHashimi:2008hr, Lee:2013lxa}, where one assumes that the gauge fields are only periodic up to a gauge transformation.

We have used our general construction to explicitly work out several examples relevant to the extraction of various EM moments, spin polarizabilities and form factors from future lattice QCD calculations. Within our construction, time-dependent magnetic fields can not be studied in this framework, which limits access to the $\gamma_{M_1M_1}$ spin polarizability of the nucleon. However, it is plausible that a more general construction will allow for this polarizability to be extracted. A rather interesting case is a plane-wave background EM field, which can be devised to be periodic on a lattice by properly choosing the Fourier modes of the fields. Through studying the corresponding QC, it is apparent that to require periodicity, either the Fourier modes in each direction must be quantized for arbitrary values of the field amplitude parameter (as implemented in Ref. \cite{Bali:2015msa}), or the Fourier modes can be chosen to be arbitrary while the amplitude parameter must be quantized. Our results have applications in upcoming lattice QCD calculations that aim to extract such quantities using the background field method.

\noindent
\subsection*{Acknowledgments}
We thank Michael G. Endres for interesting discussions and for providing the gauge configurations used in some of the numerical investigations performed in this work. We also thank Brian C. Tiburzi for fruitful correspondence. ZD was supported by the U.S. Department of Energy under grant contract number DE-SC0011090. WD was supported by the US Department of Energy Early Career Research Award DE-SC0010495.

\appendix
\section*{APPENDIX: ELECTRO/MAGNETO-PERIODIC BOUNDARY CONDITIONS AND THE ASSOCIATED CONDITIONS ON THE BACKGROUND FIELDS
\label{app:gauge-transformation}
}
\noindent
The QCs obtained in this paper for the parameters of the background fields, and the conditions that allowed a periodic implementation of $U(1)$ gauge links on the lattice, can also be deduced from imposing more general boundary conditions, namely \emph{electro/magneto-PBCs}. These boundary conditions require the gauge and matter fields to be periodic up to a gauge transformation so that all gauge-invariant quantities will be periodic. The electro/magneto-PBCs have been introduced, and extensively discussed, by 't Hooft in Refs. \cite{'tHooft19781,'tHooft1979141, 'tHooft1981} for the case of Abelian and non-Abelian gauge theories, and were later adopted to derive the QC for the case of uniform background fields implemented on a torus \cite{Smit:1986fn, Damgaard:1988hh, Rubinstein:1995hc, AlHashimi:2008hr, Bali:2011qj, Lee:2013lxa}. Here, we aim to make explicit the relation between 't Hooft's conditions and those presented in this work.\footnote{For another approach in elucidating the connection between these boundary conditions for the case of a uniform magnetic field on the lattice, see Ref. \cite{Bali:2011qj}.} In particular, we obtain those electro/magneto-PBCs that give rise to the same conditions on the $U(1)$ gauge fields that have been obtained in this paper with PBCs. It is shown that the conditions that are placed on the gauge functions when electro/magneto-PBCs are imposed (see below) are equivalent to the conditions on the flux of the EM field when PBCs are imposed. The discussions presented in this section correspond to the continuum spacetime and so can only be compared with the continuum limit of the QCs presented in Eq. (\ref{eq:QC-mu-nu-general}) (i.e., the conditions on the flux of the field strength tensor in the continuum limit). With a lattice geometry, one needs to use the results presented in earlier sections of this paper.

Consider a gauge field $A_{\mu}$ that depends on all space-time coordinates, $A_{\mu}(x_{\mu},x_{\nu},x_{\rho},x_{\sigma})$, with all indices being distinct.\footnote{The coordinate-dependence of the functions in this appendix must \emph{not} be realized through the floor-function prescription. In fact, all the functions are tilde functions as defined in the main text, but the tilde over functions is dropped to keep the notion cleaner.} For the moment, let us assume that only the $A_{\mu}$ component of the gauge field is nonzero. We demand that the field is periodic up to a gauge transformation at the boundary. Explicitly,
\begin{eqnarray}
\label{eq:Omega-I}
A_{\mu}(L_{\mu},x_{\nu},x_{\rho},x_{\sigma})&=&A_{\mu}(0,x_{\nu},x_{\rho},x_{\sigma})+\partial_{\mu}\Omega^{(\mu,\mu)}(x_{\mu},x_{\nu},x_{\rho},x_{\sigma}) ,
\\
\label{eq:Omega-II}
A_{\mu}(x_{\mu},L_{\nu},x_{\rho},x_{\sigma})&=&A_{\mu}(x_{\mu},0,x_{\rho},x_{\sigma})+\partial_{\mu}\Omega^{(\mu,\nu)}(x_{\mu},x_{\rho},x_{\sigma}),
\\
\label{eq:Omega-III}
A_{\mu}(x_{\mu},x_{\nu},L_{\rho},x_{\sigma})&=&A_{\mu}(x_{\mu},x_{\nu},0,x_{\sigma})+ \partial_{\mu}\Omega^{(\mu,\rho)}(x_{\mu},x_{\nu},x_{\sigma}),
\\
\label{eq:Omega-IV}
A_{\mu}(x_{\mu},x_{\nu},x_{\rho},L_{\sigma})&=&A_{\mu}(x_{\mu},x_{\nu},x_{\rho},0)+\partial_{\mu}\Omega^{(\mu,\sigma)}(x_{\mu},x_{\nu},x_{\rho}).
\end{eqnarray}
In order for the $\Omega$ functions to represent a gauge transformation, they must transform the matter field $\psi$ at the boundaries as well. Since the transformation of the matter fields depends on $\Omega$ (and not only the derivate of $\Omega$ with respect to $x_{\mu}$), relations (\ref{eq:Omega-I}-\ref{eq:Omega-IV}) do not entirely fix the boundary conditions on the matter fields. We now show that the following choice of electro/magneto-PBCs on matter fields,
\begin{eqnarray}
\label{eq:psi-I}
\psi(L_{\mu},x_{\nu},x_{\rho},x_{\sigma})&=&\psi(0,x_{\nu},x_{\rho},x_{\sigma}),
\\
\label{eq:psi-II}
\psi(x_{\mu},L_{\nu},x_{\rho},x_{\sigma})&=&e^{ie\hat{Q}\Omega^{(\mu,\nu)}(x_{\mu},x_{\rho},x_{\sigma})}\psi(x_{\mu},0,x_{\rho},x_{\sigma}),
\\
\label{eq:psi-III}
\psi(x_{\mu},x_{\nu},L_{\rho},x_{\sigma})&=&e^{ie\hat{Q}\Omega^{(\mu,\rho)}(x_{\mu},x_{\nu},x_{\sigma})}\psi(x_{\mu},x_{\nu},0,x_{\sigma}),
\\
\label{eq:psi-IV}
\psi(x_{\mu},x_{\nu},x_{\rho},L_{\sigma})&=&e^{ie\hat{Q}\Omega^{(\mu,\sigma)}(x_{\mu},x_{\nu},x_{\rho})}\psi(x_{\mu},x_{\nu},x_{\rho},0),
\end{eqnarray}
along with the boundary conditions on the $A_{\mu}$ fields above, gives rise to the same QCs on the background fields' flux as obtained in this paper by imposing PBCs. First note that the transformations in Eqs. (\ref{eq:Omega-I}-\ref{eq:Omega-IV}) and (\ref{eq:psi-I}-\ref{eq:psi-IV}) are consistent with the following solutions for the gauge functions,
\begin{eqnarray}
\label{eq:Omega-sol-I}
&& \Omega^{(\mu,\mu)}(x_{\mu},x_{\nu},x_{\rho},x_{\sigma})=x_{\mu}\left[ A_{\mu}(L_{\mu},x_{\nu},x_{\rho},x_{\sigma}) - A_{\mu}(0,x_{\nu},x_{\rho},x_{\sigma}) \right],
\\
\label{eq:Omega-sol-II}
&&\Omega^{(\mu,\nu)}(x_{\mu},x_{\rho},x_{\sigma})=\int_{0}^{x_\mu} dx'_{\mu} \left[ A_{\mu}(x'_{\mu},L_{\nu},x_{\rho},x_{\sigma}) - A_{\mu}(x'_{\mu},0,x_{\rho},x_{\sigma}) \right],
\\
\label{eq:Omega-sol-III}
&&\Omega^{(\mu,\rho)}(x_{\mu},x_{\nu},x_{\sigma})=\int_{0}^{x_\mu} dx'_{\mu} \left[ A_{\mu}(x'_{\mu},x_{\nu},L_{\rho},x_{\sigma})-A_{\mu}(x'_{\mu},x_{\nu},0,x_{\sigma}) \right],
\\
\label{eq:Omega-sol-IV}
&&\Omega^{(\mu,\sigma)}(x_{\mu},x_{\nu},x_{\rho})=\int_{0}^{x_\mu} dx'_{\mu} \left[ A_{\mu}(x'_{\mu},x_{\nu},x_{\rho},L_{\sigma})-A_{\mu}(x'_{\mu},x_{\nu},x_{\rho},0)\right].
\end{eqnarray}
Now consider the transformation of field $\psi$ at the corner of the lattice in the $\mu-\nu$ plane, i.e., at $x_{\mu}=L_{\mu}$ and $x_{\nu}=L_{\nu}$. One may first transform the $\psi$ field with $\Omega^{(\mu,\mu)}$ function and then with the $\Omega^{(\mu,\nu)}$ function,
\begin{eqnarray}
\psi(L_{\mu},L_{\nu},x_{\rho},x_{\sigma})=e^{ie\hat{Q}\Omega^{(\mu,\nu)}(0,x_{\rho},x_{\sigma})}\psi(0,0,x_{\rho},x_{\sigma}),
\end{eqnarray}
or the other way around,
\begin{eqnarray}
\psi(L_{\mu},L_{\nu},x_{\rho},x_{\sigma})= e^{ie\hat{Q}\Omega^{(\mu,\nu)}(L_{\mu},x_{\rho},x_{\sigma})}\psi(0,0,x_{\rho},x_{\sigma}).
\end{eqnarray}
These two relations are compatible if
\begin{eqnarray}
e^{ie\hat{Q}\left[ \Omega^{(\mu,\nu)}(0,x_{\rho},x_{\sigma}) - \Omega^{(\mu,\nu)}(L_{\mu},x_{\rho},x_{\sigma}) \right]}=1,
\label{eq:QC-omega}
\end{eqnarray}
which can only hold in general if $\Omega^{(\mu,\nu)}$ is independent of the $x_{\rho}$ and $x_{\sigma}$ coordinates. From Eqs. (\ref{eq:Omega-I}) and (\ref{eq:Omega-II}), these conditions should hold if $A_{\mu}$ is independent of these coordinates. Now given that from Eq. (\ref{eq:Omega-sol-II}), $\Omega^{(\mu,\nu)}(0,x_{\rho},x_{\sigma})=0$ and $\Omega^{(\mu,\nu)}(L_{\mu},x_{\rho},x_{\sigma})$ is the total flux of the EM field through the $\mu-\nu$ plane on the lattice (recalling the assumption that only $A_{\mu}$ is nonvanishing), the condition in Eq. (\ref{eq:QC-omega}) is identical to the continuum QC as obtained in Sec. \ref{sec:General-case},
\begin{eqnarray}
\int_{0}^{L_\mu} dx_{\mu} \left[ A_{\mu}(x_{\mu},L_{\nu}) - A_{\mu}(x_{\mu},0) \right]=\frac{2\pi n}{e\hat{Q}},~n \in \mathbb{Z}.
\label{eq:QC-flux-from-omega}
\end{eqnarray}

It might seem that by alternatively considering the gauge transformed field at the corner of the lattice in any other plane, one could arrive at the same condition which would be eliminating any space-time dependence of the gauge field. However, it is easy to see that this is not the case. For example, by considering the gauge transformed matter field in the corner of the lattice in the $\rho-\sigma$ plane one arrives at
\begin{eqnarray}
\psi(x_{\mu},x_{\nu},L_{\rho},L_{\sigma})=e^{ie\hat{Q}\Omega^{(\mu,\sigma)}(x_{\mu},x_{\nu},0)}e^{ie\hat{Q}\Omega^{(\mu,\rho)}(x_{\mu},x_{\nu},L_{\sigma})}\psi(x_{\mu},x_{\nu},0,0),
\end{eqnarray}
as well as,
\begin{eqnarray}
\psi(x_{\mu},x_{\nu},L_{\rho},L_{\sigma})=e^{ie\hat{Q}\Omega^{(\mu,\rho)}(x_{\mu},x_{\nu},0)}e^{ie\hat{Q}\Omega^{(\mu,\sigma)}(x_{\mu},x_{\nu},L_{\rho})}\psi(x_{\mu},x_{\nu},0,0),
\end{eqnarray}
but it is not hard to see from Eqs. (\ref{eq:Omega-sol-III}) and (\ref{eq:Omega-sol-IV}) that
\begin{eqnarray}
\Omega^{(\mu,\sigma)}(x_{\mu},x_{\nu},0)+\Omega^{(\mu,\rho)}(x_{\mu},x_{\nu},L_{\sigma})= \Omega^{(\mu,\rho)}(x_{\mu},x_{\nu},0)+\Omega^{(\mu,\sigma)}(x_{\mu},x_{\nu},L_{\rho}).
\end{eqnarray}
As a result, the two transformations for $\psi(x_{\mu},x_{\nu},L_{\rho},L_{\sigma})$ are identical and no extra conditions must be placed on the $x_{\mu}$ and $x_{\nu}$ dependences of the $A_{\mu}$ field.

Now let us assume that a second component of the gauge field, $A_{\nu}$, is also nonvanishing. Then the same equations as those in (\ref{eq:Omega-I}-\ref{eq:Omega-IV}) and (\ref{eq:Omega-sol-I}-\ref{eq:Omega-sol-IV}) can be written for the gauge transformed $A_{\nu}$ field at the boundaries, and for the corresponding solutions for the gauge functions, respectively. The transformations of the $\psi$ field turn into 
\begin{eqnarray}
\psi(L_{\mu},x_{\nu},x_{\rho},x_{\sigma})&=&e^{ie\hat{Q}\Omega^{(\nu,\mu)}(x_{\nu},x_{\rho},x_{\sigma})}\psi(0,x_{\nu},x_{\rho},x_{\sigma}),
\\
\psi(x_{\mu},L_{\nu},x_{\rho},x_{\sigma})&=&e^{ie\hat{Q}\Omega^{(\mu,\nu)}(x_{\mu},x_{\rho},x_{\sigma})}\psi(x_{\mu},0,x_{\rho},x_{\sigma}),
\\
\psi(x_{\mu},x_{\nu},L_{\rho},x_{\sigma})&=&e^{ie\hat{Q}\Omega^{(\nu,\rho)}(x_{\mu},x_{\nu},x_{\sigma})}e^{ie\hat{Q}\Omega^{(\mu,\rho)}(x_{\mu},x_{\nu},x_{\sigma})}\psi(x_{\mu},x_{\nu},0,x_{\sigma}),
\\
\psi(x_{\mu},x_{\nu},x_{\rho},L_{\sigma})&=&e^{ie\hat{Q}\Omega^{(\nu,\sigma)}(x_{\mu},x_{\nu},x_{\rho})}e^{ie\hat{Q}\Omega^{(\mu,\sigma)}(x_{\mu},x_{\nu},x_{\rho})}\psi(x_{\mu},x_{\nu},x_{\rho},0),
\end{eqnarray}
with our choice of boundary conditions on the matter field. The same argument discussed above based on the compatibility of the gauge transformations of the matter fields at the corner of the lattice in each plane now leads us to the condition
\begin{eqnarray}
e^{-ie\hat{Q}\left[ \Omega^{(\mu,\nu)}(0,x_{\rho},x_{\sigma}) - \Omega^{(\mu,\nu)}(L_{\mu},x_{\rho},x_{\sigma}) \right]}
e^{ie\hat{Q}\left[ \Omega^{(\nu,\mu)}(0,x_{\rho},x_{\sigma}) - \Omega^{(\nu,\mu)}(L_{\nu},x_{\rho},x_{\sigma}) \right]}=1,
\label{eq:QC-omega-mu-nu}
\end{eqnarray}
which requires that both the $A_{\mu}$ and $A_{\nu}$ fields to be independent of the $x_{\rho}$ and $x_{\sigma}$ coordinates. Additionally, it is easy to see that this condition implies that the total flux of the EM field through the $\mu-\nu$ plane is quantized,
\begin{eqnarray}
\int_{0}^{L_\mu} dx_{\mu} \left[ A_{\mu}(x_{\mu},L_{\nu}) - A_{\mu}(x_{\mu},0) \right]-\int_{0}^{L_\nu} dx_{\nu} \left[ A_{\nu}(L_{\mu},x_{\nu}) - A_{\nu}(0,x_{\nu}) \right]=\frac{2\pi n}{e\hat{Q}},~n \in \mathbb{Z},
\label{eq:QC-omega-mu-nu}
\end{eqnarray}
conditions that were all deduced in Sec. \ref{sec:General-case} by imposing PBCs. In general, if any of the $A_{\mu}$ component of the gauge field with $\mu=0,1,2,3$ is nonvanishing, the compatibility relations similar to the ones considered above must be studied at the corner of the lattice in all planes, and they are all satisfied if the flux of the field strength tensor through each plane is coordinate-independent and quantized.  

We conclude this appendix by a remark. Note that the boundary conditions in this section are defined by only identifying the point $L_{\mu}$ with the point $0$ in the $\mu$ direction, and similarly in other directions. If one imposes more restrictive conditions such that all points $x_{\mu}+L_{\mu}$ are identified with points $x_{\mu}$, the QCs derived in this section for the flux of the background field through the $\mu-\nu$ plane will depend, in general, on the $x_{\mu}$ and $x_{\nu}$ coordinates and can not be quantized. This is in contrast with the case of uniform background fields where both the PBCs and electro/magneto-PBCs can be satisfied with a single space/time-independent QC. Such system possesses a translational invariance in units of $L_{\mu}$ for all $\mu$ (corresponding to the magnetic translational group in the case of a uniform magnetic field, see Ref. \cite{AlHashimi:2008hr}). With a background EM field that is space-time dependent, no translational invariance exists prior to imposing the boundary conditions. Therefore, the lack of discrete translational invariance in units of $L_{\mu}$ should not come as a surprise. Given our choice of periodifying the functions with the use of the floor function, we have however explicitly built up such translational invariance in the setup presented in this paper. One may however wonder whether this choice can be distinguished from a choice for which only the point $L_{\mu}$ is identified with the point $0$ on the lattice. This answer is that with a lattice action that at most depends on the first derivative of the gauge and matter fields, these two choices are identical at the practical level. Explicitly, one only evaluates fields at points $0 \leq x_{\mu}  < L_{\mu}$, and only the values of the fields at $L_{\mu}$ must be specified through the boundary conditions.

\bibliography{bibi}

\begin{thebibliography}{53}
\expandafter\ifx\csname natexlab\endcsname\relax\def\natexlab#1{#1}\fi
\expandafter\ifx\csname bibnamefont\endcsname\relax
  \def\bibnamefont#1{#1}\fi
\expandafter\ifx\csname bibfnamefont\endcsname\relax
  \def\bibfnamefont#1{#1}\fi
\expandafter\ifx\csname citenamefont\endcsname\relax
  \def\citenamefont#1{#1}\fi
\expandafter\ifx\csname url\endcsname\relax
  \def\url#1{\texttt{#1}}\fi
\expandafter\ifx\csname urlprefix\endcsname\relax\def\urlprefix{URL }\fi
\providecommand{\bibinfo}[2]{#2}
\providecommand{\eprint}[2][]{\url{#2}}

\bibitem[{\citenamefont{Blum et~al.}(2007)\citenamefont{Blum, Doi, Hayakawa,
  Izubuchi, and Yamada}}]{Blum:2007cy}
\bibinfo{author}{\bibfnamefont{T.}~\bibnamefont{Blum}},
  \bibinfo{author}{\bibfnamefont{T.}~\bibnamefont{Doi}},
  \bibinfo{author}{\bibfnamefont{M.}~\bibnamefont{Hayakawa}},
  \bibinfo{author}{\bibfnamefont{T.}~\bibnamefont{Izubuchi}}, \bibnamefont{and}
  \bibinfo{author}{\bibfnamefont{N.}~\bibnamefont{Yamada}},
  \bibinfo{journal}{Phys.Rev.} \textbf{\bibinfo{volume}{D76}},
  \bibinfo{pages}{114508} (\bibinfo{year}{2007}), \eprint{0708.0484}.

\bibitem[{\citenamefont{Basak et~al.}(2008)}]{Basak:2008na}
\bibinfo{author}{\bibfnamefont{S.}~\bibnamefont{Basak}} \bibnamefont{et~al.}
  (\bibinfo{collaboration}{MILC Collaboration}), \bibinfo{journal}{PoS}
  \textbf{\bibinfo{volume}{LATTICE2008}}, \bibinfo{pages}{127}
  (\bibinfo{year}{2008}), \eprint{0812.4486}.

\bibitem[{\citenamefont{Blum et~al.}(2010)\citenamefont{Blum, Zhou, Doi,
  Hayakawa, Izubuchi et~al.}}]{Blum:2010ym}
\bibinfo{author}{\bibfnamefont{T.}~\bibnamefont{Blum}},
  \bibinfo{author}{\bibfnamefont{R.}~\bibnamefont{Zhou}},
  \bibinfo{author}{\bibfnamefont{T.}~\bibnamefont{Doi}},
  \bibinfo{author}{\bibfnamefont{M.}~\bibnamefont{Hayakawa}},
  \bibinfo{author}{\bibfnamefont{T.}~\bibnamefont{Izubuchi}},
  \bibnamefont{et~al.}, \bibinfo{journal}{Phys.Rev.}
  \textbf{\bibinfo{volume}{D82}}, \bibinfo{pages}{094508}
  (\bibinfo{year}{2010}), \eprint{1006.1311}.

\bibitem[{\citenamefont{Portelli et~al.}(2010)}]{Portelli:2010yn}
\bibinfo{author}{\bibfnamefont{A.}~\bibnamefont{Portelli}} \bibnamefont{et~al.}
  (\bibinfo{collaboration}{Budapest-Marseille-Wuppertal Collaboration}),
  \bibinfo{journal}{PoS} \textbf{\bibinfo{volume}{LATTICE2010}},
  \bibinfo{pages}{121} (\bibinfo{year}{2010}), \eprint{1011.4189}.

\bibitem[{\citenamefont{Portelli et~al.}(2011)\citenamefont{Portelli, Durr,
  Fodor, Frison, Hoelbling et~al.}}]{Portelli:2012pn}
\bibinfo{author}{\bibfnamefont{A.}~\bibnamefont{Portelli}},
  \bibinfo{author}{\bibfnamefont{S.}~\bibnamefont{Durr}},
  \bibinfo{author}{\bibfnamefont{Z.}~\bibnamefont{Fodor}},
  \bibinfo{author}{\bibfnamefont{J.}~\bibnamefont{Frison}},
  \bibinfo{author}{\bibfnamefont{C.}~\bibnamefont{Hoelbling}},
  \bibnamefont{et~al.}, \bibinfo{journal}{PoS}
  \textbf{\bibinfo{volume}{LATTICE2011}}, \bibinfo{pages}{136}
  (\bibinfo{year}{2011}), \eprint{1201.2787}.

\bibitem[{\citenamefont{Aoki et~al.}(2012)\citenamefont{Aoki, Ishikawa,
  Ishizuka, Kanaya, Kuramashi et~al.}}]{Aoki:2012st}
\bibinfo{author}{\bibfnamefont{S.}~\bibnamefont{Aoki}},
  \bibinfo{author}{\bibfnamefont{K.}~\bibnamefont{Ishikawa}},
  \bibinfo{author}{\bibfnamefont{N.}~\bibnamefont{Ishizuka}},
  \bibinfo{author}{\bibfnamefont{K.}~\bibnamefont{Kanaya}},
  \bibinfo{author}{\bibfnamefont{Y.}~\bibnamefont{Kuramashi}},
  \bibnamefont{et~al.}, \bibinfo{journal}{Phys.Rev.}
  \textbf{\bibinfo{volume}{D86}}, \bibinfo{pages}{034507}
  (\bibinfo{year}{2012}), \eprint{1205.2961}.

\bibitem[{\citenamefont{de~Divitiis et~al.}(2013)\citenamefont{de~Divitiis,
  Frezzotti, Lubicz, Martinelli, Petronzio et~al.}}]{deDivitiis:2013xla}
\bibinfo{author}{\bibfnamefont{G.}~\bibnamefont{de~Divitiis}},
  \bibinfo{author}{\bibfnamefont{R.}~\bibnamefont{Frezzotti}},
  \bibinfo{author}{\bibfnamefont{V.}~\bibnamefont{Lubicz}},
  \bibinfo{author}{\bibfnamefont{G.}~\bibnamefont{Martinelli}},
  \bibinfo{author}{\bibfnamefont{R.}~\bibnamefont{Petronzio}},
  \bibnamefont{et~al.}, \bibinfo{journal}{Phys.Rev.}
  \textbf{\bibinfo{volume}{D87}}, \bibinfo{pages}{114505}
  (\bibinfo{year}{2013}), \eprint{1303.4896}.

\bibitem[{\citenamefont{Borsanyi et~al.}(2013)\citenamefont{Borsanyi, Durr,
  Fodor, Frison, Hoelbling et~al.}}]{Borsanyi:2013lga}
\bibinfo{author}{\bibfnamefont{S.}~\bibnamefont{Borsanyi}},
  \bibinfo{author}{\bibfnamefont{S.}~\bibnamefont{Durr}},
  \bibinfo{author}{\bibfnamefont{Z.}~\bibnamefont{Fodor}},
  \bibinfo{author}{\bibfnamefont{J.}~\bibnamefont{Frison}},
  \bibinfo{author}{\bibfnamefont{C.}~\bibnamefont{Hoelbling}},
  \bibnamefont{et~al.}, \bibinfo{journal}{Phys.Rev.Lett.}
  \textbf{\bibinfo{volume}{111}}, \bibinfo{pages}{252001}
  (\bibinfo{year}{2013}), \eprint{1306.2287}.

\bibitem[{\citenamefont{Drury et~al.}(2013)\citenamefont{Drury, Blum, Hayakawa,
  Izubuchi, Sachrajda et~al.}}]{Drury:2013sfa}
\bibinfo{author}{\bibfnamefont{S.}~\bibnamefont{Drury}},
  \bibinfo{author}{\bibfnamefont{T.}~\bibnamefont{Blum}},
  \bibinfo{author}{\bibfnamefont{M.}~\bibnamefont{Hayakawa}},
  \bibinfo{author}{\bibfnamefont{T.}~\bibnamefont{Izubuchi}},
  \bibinfo{author}{\bibfnamefont{C.}~\bibnamefont{Sachrajda}},
  \bibnamefont{et~al.} (\bibinfo{year}{2013}), \eprint{1312.0477}.

\bibitem[{\citenamefont{Borsanyi et~al.}(2014)\citenamefont{Borsanyi, Durr,
  Fodor, Hoelbling, Katz et~al.}}]{Borsanyi:2014jba}
\bibinfo{author}{\bibfnamefont{S.}~\bibnamefont{Borsanyi}},
  \bibinfo{author}{\bibfnamefont{S.}~\bibnamefont{Durr}},
  \bibinfo{author}{\bibfnamefont{Z.}~\bibnamefont{Fodor}},
  \bibinfo{author}{\bibfnamefont{C.}~\bibnamefont{Hoelbling}},
  \bibinfo{author}{\bibfnamefont{S.}~\bibnamefont{Katz}}, \bibnamefont{et~al.}
  (\bibinfo{year}{2014}), \eprint{1406.4088}.

\bibitem[{\citenamefont{Blum et~al.}(2015)\citenamefont{Blum, Chowdhury,
  Hayakawa, and Izubuchi}}]{Blum:2014oka}
\bibinfo{author}{\bibfnamefont{T.}~\bibnamefont{Blum}},
  \bibinfo{author}{\bibfnamefont{S.}~\bibnamefont{Chowdhury}},
  \bibinfo{author}{\bibfnamefont{M.}~\bibnamefont{Hayakawa}}, \bibnamefont{and}
  \bibinfo{author}{\bibfnamefont{T.}~\bibnamefont{Izubuchi}},
  \bibinfo{journal}{Phys.Rev.Lett.} \textbf{\bibinfo{volume}{114}},
  \bibinfo{pages}{012001} (\bibinfo{year}{2015}), \eprint{1407.2923}.

\bibitem[{\citenamefont{Hagler}(2010)}]{Hagler:2009ni}
\bibinfo{author}{\bibfnamefont{P.}~\bibnamefont{Hagler}},
  \bibinfo{journal}{Phys.Rept.} \textbf{\bibinfo{volume}{490}},
  \bibinfo{pages}{49} (\bibinfo{year}{2010}), \eprint{0912.5483}.

\bibitem[{\citenamefont{Lee and Tiburzi}(2014{\natexlab{a}})}]{Lee:2013lxa}
\bibinfo{author}{\bibfnamefont{J.-W.} \bibnamefont{Lee}} \bibnamefont{and}
  \bibinfo{author}{\bibfnamefont{B.~C.} \bibnamefont{Tiburzi}},
  \bibinfo{journal}{Phys.Rev.} \textbf{\bibinfo{volume}{D89}},
  \bibinfo{pages}{054017} (\bibinfo{year}{2014}{\natexlab{a}}),
  \eprint{1312.3969}.

\bibitem[{\citenamefont{Lee and Tiburzi}(2014{\natexlab{b}})}]{Lee:2014iha}
\bibinfo{author}{\bibfnamefont{J.-W.} \bibnamefont{Lee}} \bibnamefont{and}
  \bibinfo{author}{\bibfnamefont{B.~C.} \bibnamefont{Tiburzi}},
  \bibinfo{journal}{Phys.Rev.} \textbf{\bibinfo{volume}{D90}},
  \bibinfo{pages}{074036} (\bibinfo{year}{2014}{\natexlab{b}}),
  \eprint{1407.8159}.

\bibitem[{\citenamefont{Davoudi and Detmold}(2015)}]{davoudi2015}
\bibinfo{author}{\bibfnamefont{Z.}~\bibnamefont{Davoudi}} \bibnamefont{and}
  \bibinfo{author}{\bibfnamefont{W.}~\bibnamefont{Detmold}}
  (\bibinfo{year}{2015}), \bibinfo{note}{under preparation.}

\bibitem[{\citenamefont{Caswell and Lepage}(1986)}]{Caswell:1985ui}
\bibinfo{author}{\bibfnamefont{W.}~\bibnamefont{Caswell}} \bibnamefont{and}
  \bibinfo{author}{\bibfnamefont{G.}~\bibnamefont{Lepage}},
  \bibinfo{journal}{Phys.Lett.} \textbf{\bibinfo{volume}{B167}},
  \bibinfo{pages}{437} (\bibinfo{year}{1986}).

\bibitem[{\citenamefont{Labelle}(1992)}]{Labelle:1992hd}
\bibinfo{author}{\bibfnamefont{P.}~\bibnamefont{Labelle}}
  (\bibinfo{year}{1992}), \eprint{9209266}.

\bibitem[{\citenamefont{Labelle et~al.}(1997)\citenamefont{Labelle, Zebarjad,
  and Burgess}}]{Labelle:1997uw}
\bibinfo{author}{\bibfnamefont{P.}~\bibnamefont{Labelle}},
  \bibinfo{author}{\bibfnamefont{S.}~\bibnamefont{Zebarjad}}, \bibnamefont{and}
  \bibinfo{author}{\bibfnamefont{C.}~\bibnamefont{Burgess}},
  \bibinfo{journal}{Phys.Rev.} \textbf{\bibinfo{volume}{D56}},
  \bibinfo{pages}{8053} (\bibinfo{year}{1997}), \eprint{9706449}.

\bibitem[{\citenamefont{Kaplan et~al.}(1998{\natexlab{a}})\citenamefont{Kaplan,
  Savage, and Wise}}]{Kaplan:1998tg}
\bibinfo{author}{\bibfnamefont{D.~B.} \bibnamefont{Kaplan}},
  \bibinfo{author}{\bibfnamefont{M.~J.} \bibnamefont{Savage}},
  \bibnamefont{and} \bibinfo{author}{\bibfnamefont{M.~B.} \bibnamefont{Wise}},
  \bibinfo{journal}{Phys.Lett.} \textbf{\bibinfo{volume}{B424}},
  \bibinfo{pages}{390} (\bibinfo{year}{1998}{\natexlab{a}}), \eprint{9801034}.

\bibitem[{\citenamefont{Kaplan et~al.}(1998{\natexlab{b}})\citenamefont{Kaplan,
  Savage, and Wise}}]{Kaplan:1998we}
\bibinfo{author}{\bibfnamefont{D.~B.} \bibnamefont{Kaplan}},
  \bibinfo{author}{\bibfnamefont{M.~J.} \bibnamefont{Savage}},
  \bibnamefont{and} \bibinfo{author}{\bibfnamefont{M.~B.} \bibnamefont{Wise}},
  \bibinfo{journal}{Nucl.Phys.} \textbf{\bibinfo{volume}{B534}},
  \bibinfo{pages}{329} (\bibinfo{year}{1998}{\natexlab{b}}), \eprint{9802075}.

\bibitem[{\citenamefont{Chen et~al.}(1999)\citenamefont{Chen, Rupak, and
  Savage}}]{Chen:1999tn}
\bibinfo{author}{\bibfnamefont{J.-W.} \bibnamefont{Chen}},
  \bibinfo{author}{\bibfnamefont{G.}~\bibnamefont{Rupak}}, \bibnamefont{and}
  \bibinfo{author}{\bibfnamefont{M.~J.} \bibnamefont{Savage}},
  \bibinfo{journal}{Nucl.Phys.} \textbf{\bibinfo{volume}{A653}},
  \bibinfo{pages}{386} (\bibinfo{year}{1999}), \eprint{9902056}.

\bibitem[{\citenamefont{Detmold et~al.}(2006)\citenamefont{Detmold, Tiburzi,
  and Walker-Loud}}]{Detmold:2006vu}
\bibinfo{author}{\bibfnamefont{W.}~\bibnamefont{Detmold}},
  \bibinfo{author}{\bibfnamefont{B.}~\bibnamefont{Tiburzi}}, \bibnamefont{and}
  \bibinfo{author}{\bibfnamefont{A.}~\bibnamefont{Walker-Loud}},
  \bibinfo{journal}{Phys.Rev.} \textbf{\bibinfo{volume}{D73}},
  \bibinfo{pages}{114505} (\bibinfo{year}{2006}), \eprint{0603026}.

\bibitem[{\citenamefont{Detmold et~al.}(2009)\citenamefont{Detmold, Tiburzi,
  and Walker-Loud}}]{Detmold:2009dx}
\bibinfo{author}{\bibfnamefont{W.}~\bibnamefont{Detmold}},
  \bibinfo{author}{\bibfnamefont{B.~C.} \bibnamefont{Tiburzi}},
  \bibnamefont{and}
  \bibinfo{author}{\bibfnamefont{A.}~\bibnamefont{Walker-Loud}},
  \bibinfo{journal}{Phys.Rev.} \textbf{\bibinfo{volume}{D79}},
  \bibinfo{pages}{094505} (\bibinfo{year}{2009}), \eprint{0904.1586}.

\bibitem[{\citenamefont{Hill et~al.}(2013)\citenamefont{Hill, Lee, Paz, and
  Solon}}]{Hill:2012rh}
\bibinfo{author}{\bibfnamefont{R.~J.} \bibnamefont{Hill}},
  \bibinfo{author}{\bibfnamefont{G.}~\bibnamefont{Lee}},
  \bibinfo{author}{\bibfnamefont{G.}~\bibnamefont{Paz}}, \bibnamefont{and}
  \bibinfo{author}{\bibfnamefont{M.~P.} \bibnamefont{Solon}},
  \bibinfo{journal}{Phys.Rev.} \textbf{\bibinfo{volume}{D87}},
  \bibinfo{pages}{053017} (\bibinfo{year}{2013}), \eprint{1212.4508}.

\bibitem[{\citenamefont{Bernard et~al.}(1982)\citenamefont{Bernard, Draper,
  Olynyk, and Rushton}}]{Bernard:1982yu}
\bibinfo{author}{\bibfnamefont{C.~W.} \bibnamefont{Bernard}},
  \bibinfo{author}{\bibfnamefont{T.}~\bibnamefont{Draper}},
  \bibinfo{author}{\bibfnamefont{K.}~\bibnamefont{Olynyk}}, \bibnamefont{and}
  \bibinfo{author}{\bibfnamefont{M.}~\bibnamefont{Rushton}},
  \bibinfo{journal}{Phys.Rev.Lett.} \textbf{\bibinfo{volume}{49}},
  \bibinfo{pages}{1076} (\bibinfo{year}{1982}).

\bibitem[{\citenamefont{Martinelli et~al.}(1982)\citenamefont{Martinelli,
  Parisi, Petronzio, and Rapuano}}]{Martinelli:1982cb}
\bibinfo{author}{\bibfnamefont{G.}~\bibnamefont{Martinelli}},
  \bibinfo{author}{\bibfnamefont{G.}~\bibnamefont{Parisi}},
  \bibinfo{author}{\bibfnamefont{R.}~\bibnamefont{Petronzio}},
  \bibnamefont{and} \bibinfo{author}{\bibfnamefont{F.}~\bibnamefont{Rapuano}},
  \bibinfo{journal}{Phys.Lett.} \textbf{\bibinfo{volume}{B116}},
  \bibinfo{pages}{434} (\bibinfo{year}{1982}).

\bibitem[{\citenamefont{Fiebig et~al.}(1989)\citenamefont{Fiebig, Wilcox, and
  Woloshyn}}]{Fiebig:1988en}
\bibinfo{author}{\bibfnamefont{H.}~\bibnamefont{Fiebig}},
  \bibinfo{author}{\bibfnamefont{W.}~\bibnamefont{Wilcox}}, \bibnamefont{and}
  \bibinfo{author}{\bibfnamefont{R.}~\bibnamefont{Woloshyn}},
  \bibinfo{journal}{Nucl.Phys.} \textbf{\bibinfo{volume}{B324}},
  \bibinfo{pages}{47} (\bibinfo{year}{1989}).

\bibitem[{\citenamefont{Christensen et~al.}(2005)\citenamefont{Christensen,
  Wilcox, Lee, and Zhou}}]{Christensen:2004ca}
\bibinfo{author}{\bibfnamefont{J.~C.} \bibnamefont{Christensen}},
  \bibinfo{author}{\bibfnamefont{W.}~\bibnamefont{Wilcox}},
  \bibinfo{author}{\bibfnamefont{F.~X.} \bibnamefont{Lee}}, \bibnamefont{and}
  \bibinfo{author}{\bibfnamefont{L.}~\bibnamefont{Zhou}},
  \bibinfo{journal}{Phys.Rev.} \textbf{\bibinfo{volume}{D72}},
  \bibinfo{pages}{034503} (\bibinfo{year}{2005}), \eprint{0408024}.

\bibitem[{\citenamefont{Lee et~al.}(2005)\citenamefont{Lee, Kelly, Zhou, and
  Wilcox}}]{Lee:2005ds}
\bibinfo{author}{\bibfnamefont{F.}~\bibnamefont{Lee}},
  \bibinfo{author}{\bibfnamefont{R.}~\bibnamefont{Kelly}},
  \bibinfo{author}{\bibfnamefont{L.}~\bibnamefont{Zhou}}, \bibnamefont{and}
  \bibinfo{author}{\bibfnamefont{W.}~\bibnamefont{Wilcox}},
  \bibinfo{journal}{Phys.Lett.} \textbf{\bibinfo{volume}{B627}},
  \bibinfo{pages}{71} (\bibinfo{year}{2005}), \eprint{0509067}.

\bibitem[{\citenamefont{Lee et~al.}(2006)\citenamefont{Lee, Zhou, Wilcox, and
  Christensen}}]{Lee:2005dq}
\bibinfo{author}{\bibfnamefont{F.~X.} \bibnamefont{Lee}},
  \bibinfo{author}{\bibfnamefont{L.}~\bibnamefont{Zhou}},
  \bibinfo{author}{\bibfnamefont{W.}~\bibnamefont{Wilcox}}, \bibnamefont{and}
  \bibinfo{author}{\bibfnamefont{J.~C.} \bibnamefont{Christensen}},
  \bibinfo{journal}{Phys.Rev.} \textbf{\bibinfo{volume}{D73}},
  \bibinfo{pages}{034503} (\bibinfo{year}{2006}), \eprint{0509065}.

\bibitem[{\citenamefont{Aubin et~al.}(2009)\citenamefont{Aubin, Orginos,
  Pascalutsa, and Vanderhaeghen}}]{Aubin:2008qp}
\bibinfo{author}{\bibfnamefont{C.}~\bibnamefont{Aubin}},
  \bibinfo{author}{\bibfnamefont{K.}~\bibnamefont{Orginos}},
  \bibinfo{author}{\bibfnamefont{V.}~\bibnamefont{Pascalutsa}},
  \bibnamefont{and}
  \bibinfo{author}{\bibfnamefont{M.}~\bibnamefont{Vanderhaeghen}},
  \bibinfo{journal}{Phys.Rev.} \textbf{\bibinfo{volume}{D79}},
  \bibinfo{pages}{051502} (\bibinfo{year}{2009}), \eprint{0811.2440}.

\bibitem[{\citenamefont{Detmold et~al.}(2010)\citenamefont{Detmold, Tiburzi,
  and Walker-Loud}}]{Detmold:2010ts}
\bibinfo{author}{\bibfnamefont{W.}~\bibnamefont{Detmold}},
  \bibinfo{author}{\bibfnamefont{B.}~\bibnamefont{Tiburzi}}, \bibnamefont{and}
  \bibinfo{author}{\bibfnamefont{A.}~\bibnamefont{Walker-Loud}},
  \bibinfo{journal}{Phys.Rev.} \textbf{\bibinfo{volume}{D81}},
  \bibinfo{pages}{054502} (\bibinfo{year}{2010}), \eprint{1001.1131}.

\bibitem[{\citenamefont{Primer et~al.}(2014)\citenamefont{Primer, Kamleh,
  Leinweber, and Burkardt}}]{Primer:2013pva}
\bibinfo{author}{\bibfnamefont{T.}~\bibnamefont{Primer}},
  \bibinfo{author}{\bibfnamefont{W.}~\bibnamefont{Kamleh}},
  \bibinfo{author}{\bibfnamefont{D.}~\bibnamefont{Leinweber}},
  \bibnamefont{and} \bibinfo{author}{\bibfnamefont{M.}~\bibnamefont{Burkardt}},
  \bibinfo{journal}{Phys.Rev.} \textbf{\bibinfo{volume}{D89}},
  \bibinfo{pages}{034508} (\bibinfo{year}{2014}), \eprint{1307.1509}.

\bibitem[{\citenamefont{Lujan et~al.}(2014)\citenamefont{Lujan, Alexandru,
  Freeman, and Lee}}]{Lujan:2014kia}
\bibinfo{author}{\bibfnamefont{M.}~\bibnamefont{Lujan}},
  \bibinfo{author}{\bibfnamefont{A.}~\bibnamefont{Alexandru}},
  \bibinfo{author}{\bibfnamefont{W.}~\bibnamefont{Freeman}}, \bibnamefont{and}
  \bibinfo{author}{\bibfnamefont{F.}~\bibnamefont{Lee}},
  \bibinfo{journal}{Phys.Rev.} \textbf{\bibinfo{volume}{D89}},
  \bibinfo{pages}{074506} (\bibinfo{year}{2014}), \eprint{1402.3025}.

\bibitem[{\citenamefont{Beane et~al.}(2014)\citenamefont{Beane, Chang, Cohen,
  Detmold, Lin, Orginos, Parreno, Savage, and Tiburzi}}]{Beane:2014ora}
\bibinfo{author}{\bibfnamefont{S.~R.} \bibnamefont{Beane}},
  \bibinfo{author}{\bibfnamefont{E.}~\bibnamefont{Chang}},
  \bibinfo{author}{\bibfnamefont{S.}~\bibnamefont{Cohen}},
  \bibinfo{author}{\bibfnamefont{W.}~\bibnamefont{Detmold}},
  \bibinfo{author}{\bibfnamefont{H.}~\bibnamefont{Lin}},
  \bibinfo{author}{\bibfnamefont{K.}~\bibnamefont{Orginos}},
  \bibinfo{author}{\bibfnamefont{A.}~\bibnamefont{Parreno}},
  \bibinfo{author}{\bibfnamefont{M.~J.} \bibnamefont{Savage}},
  \bibnamefont{and} \bibinfo{author}{\bibfnamefont{B.~C.}
  \bibnamefont{Tiburzi}}, \bibinfo{journal}{Phys.Rev.Lett.}
  \textbf{\bibinfo{volume}{113}}, \bibinfo{pages}{252001}
  (\bibinfo{year}{2014}), \eprint{1409.3556}.

\bibitem[{\citenamefont{Beane et~al.}(2015)\citenamefont{Beane, Chang, Detmold,
  Orginos, Parreño et~al.}}]{Beane:2015yha}
\bibinfo{author}{\bibfnamefont{S.~R.} \bibnamefont{Beane}},
  \bibinfo{author}{\bibfnamefont{E.}~\bibnamefont{Chang}},
  \bibinfo{author}{\bibfnamefont{W.}~\bibnamefont{Detmold}},
  \bibinfo{author}{\bibfnamefont{K.}~\bibnamefont{Orginos}},
  \bibinfo{author}{\bibfnamefont{A.}~\bibnamefont{Parreño}},
  \bibnamefont{et~al.} (\bibinfo{year}{2015}), \eprint{1505.02422}.

\bibitem[{\citenamefont{Chang et~al.}(2015)\citenamefont{Chang, Detmold,
  Orginos, Parreno, Savage et~al.}}]{Chang:2015qxa}
\bibinfo{author}{\bibfnamefont{E.}~\bibnamefont{Chang}},
  \bibinfo{author}{\bibfnamefont{W.}~\bibnamefont{Detmold}},
  \bibinfo{author}{\bibfnamefont{K.}~\bibnamefont{Orginos}},
  \bibinfo{author}{\bibfnamefont{A.}~\bibnamefont{Parreno}},
  \bibinfo{author}{\bibfnamefont{M.~J.} \bibnamefont{Savage}},
  \bibnamefont{et~al.} (\bibinfo{year}{2015}), \eprint{1506.05518}.

\bibitem[{\citenamefont{Ragusa}(1993)}]{PhysRevD.47.3757}
\bibinfo{author}{\bibfnamefont{S.}~\bibnamefont{Ragusa}},
  \bibinfo{journal}{Phys. Rev. D} \textbf{\bibinfo{volume}{47}},
  \bibinfo{pages}{3757} (\bibinfo{year}{1993}).

\bibitem[{\citenamefont{Detmold}(2005)}]{Detmold:2004kw}
\bibinfo{author}{\bibfnamefont{W.}~\bibnamefont{Detmold}},
  \bibinfo{journal}{Phys.Rev.} \textbf{\bibinfo{volume}{D71}},
  \bibinfo{pages}{054506} (\bibinfo{year}{2005}), \eprint{0410011}.

\bibitem[{\citenamefont{Bali and Endrodi}(2015)}]{Bali:2015msa}
\bibinfo{author}{\bibfnamefont{G.}~\bibnamefont{Bali}} \bibnamefont{and}
  \bibinfo{author}{\bibfnamefont{G.}~\bibnamefont{Endrodi}}
  (\bibinfo{year}{2015}), \eprint{1506.08638}.

\bibitem[{\citenamefont{Lee and Alexandru}(2011)}]{Lee:2011gz}
\bibinfo{author}{\bibfnamefont{F.~X.} \bibnamefont{Lee}} \bibnamefont{and}
  \bibinfo{author}{\bibfnamefont{A.}~\bibnamefont{Alexandru}},
  \bibinfo{journal}{PoS} \textbf{\bibinfo{volume}{LATTICE2011}},
  \bibinfo{pages}{317} (\bibinfo{year}{2011}), \eprint{1111.4425}.

\bibitem[{\citenamefont{Engelhardt}(2011)}]{Engelhardt:2011qq}
\bibinfo{author}{\bibfnamefont{M.}~\bibnamefont{Engelhardt}},
  \bibinfo{journal}{PoS} \textbf{\bibinfo{volume}{LATTICE2011}},
  \bibinfo{pages}{153} (\bibinfo{year}{2011}), \eprint{1111.3686}.

\bibitem[{\citenamefont{Detmold et~al.}(2008)\citenamefont{Detmold, Tiburzi,
  and Walker-Loud}}]{Detmold:2008xk}
\bibinfo{author}{\bibfnamefont{W.}~\bibnamefont{Detmold}},
  \bibinfo{author}{\bibfnamefont{B.~C.} \bibnamefont{Tiburzi}},
  \bibnamefont{and}
  \bibinfo{author}{\bibfnamefont{A.}~\bibnamefont{Walker-Loud}},
  \bibinfo{journal}{PoS} \textbf{\bibinfo{volume}{LATTICE2008}},
  \bibinfo{pages}{147} (\bibinfo{year}{2008}), \eprint{0809.0721}.

\bibitem[{\citenamefont{Tiburzi}(2013)}]{Tiburzi:2013vza}
\bibinfo{author}{\bibfnamefont{B.}~\bibnamefont{Tiburzi}},
  \bibinfo{journal}{Phys.Rev.} \textbf{\bibinfo{volume}{D88}},
  \bibinfo{pages}{034027} (\bibinfo{year}{2013}), \eprint{1302.6645}.

\bibitem[{\citenamefont{'t~Hooft}(1979)}]{'tHooft1979141}
\bibinfo{author}{\bibfnamefont{G.}~\bibnamefont{'t~Hooft}},
  \bibinfo{journal}{Nuclear Physics B} \textbf{\bibinfo{volume}{153}},
  \bibinfo{pages}{141 } (\bibinfo{year}{1979}), ISSN \bibinfo{issn}{0550-3213}.

\bibitem[{\citenamefont{Smit and Vink}(1987)}]{Smit:1986fn}
\bibinfo{author}{\bibfnamefont{J.}~\bibnamefont{Smit}} \bibnamefont{and}
  \bibinfo{author}{\bibfnamefont{J.~C.} \bibnamefont{Vink}},
  \bibinfo{journal}{Nucl.Phys.} \textbf{\bibinfo{volume}{B286}},
  \bibinfo{pages}{485} (\bibinfo{year}{1987}).

\bibitem[{\citenamefont{Al-Hashimi and Wiese}(2009)}]{AlHashimi:2008hr}
\bibinfo{author}{\bibfnamefont{M.}~\bibnamefont{Al-Hashimi}} \bibnamefont{and}
  \bibinfo{author}{\bibfnamefont{U.-J.} \bibnamefont{Wiese}},
  \bibinfo{journal}{Annals Phys.} \textbf{\bibinfo{volume}{324}},
  \bibinfo{pages}{343} (\bibinfo{year}{2009}), \eprint{0807.0630}.

\bibitem[{\citenamefont{Bali et~al.}(2012)\citenamefont{Bali, Bruckmann,
  Endrodi, Fodor, Katz et~al.}}]{Bali:2011qj}
\bibinfo{author}{\bibfnamefont{G.}~\bibnamefont{Bali}},
  \bibinfo{author}{\bibfnamefont{F.}~\bibnamefont{Bruckmann}},
  \bibinfo{author}{\bibfnamefont{G.}~\bibnamefont{Endrodi}},
  \bibinfo{author}{\bibfnamefont{Z.}~\bibnamefont{Fodor}},
  \bibinfo{author}{\bibfnamefont{S.}~\bibnamefont{Katz}}, \bibnamefont{et~al.},
  \bibinfo{journal}{JHEP} \textbf{\bibinfo{volume}{1202}}, \bibinfo{pages}{044}
  (\bibinfo{year}{2012}), \eprint{1111.4956}.

\bibitem[{\citenamefont{'t~Hooft}(1978)}]{'tHooft19781}
\bibinfo{author}{\bibfnamefont{G.}~\bibnamefont{'t~Hooft}},
  \bibinfo{journal}{Nuclear Physics B} \textbf{\bibinfo{volume}{138}},
  \bibinfo{pages}{1 } (\bibinfo{year}{1978}), ISSN \bibinfo{issn}{0550-3213}.

\bibitem[{\citenamefont{'t~Hooft}(1981)}]{'tHooft1981}
\bibinfo{author}{\bibfnamefont{G.}~\bibnamefont{'t~Hooft}},
  \bibinfo{journal}{Communications in Mathematical Physics}
  \textbf{\bibinfo{volume}{81}} (\bibinfo{year}{1981}), ISSN
  \bibinfo{issn}{0010-3616}.

\bibitem[{\citenamefont{Damgaard and Heller}(1988)}]{Damgaard:1988hh}
\bibinfo{author}{\bibfnamefont{P.~H.} \bibnamefont{Damgaard}} \bibnamefont{and}
  \bibinfo{author}{\bibfnamefont{U.~M.} \bibnamefont{Heller}},
  \bibinfo{journal}{Nucl.Phys.} \textbf{\bibinfo{volume}{B309}},
  \bibinfo{pages}{625} (\bibinfo{year}{1988}).

\bibitem[{\citenamefont{Rubinstein et~al.}(1995)\citenamefont{Rubinstein,
  Solomon, and Wittlich}}]{Rubinstein:1995hc}
\bibinfo{author}{\bibfnamefont{H.}~\bibnamefont{Rubinstein}},
  \bibinfo{author}{\bibfnamefont{S.}~\bibnamefont{Solomon}}, \bibnamefont{and}
  \bibinfo{author}{\bibfnamefont{T.}~\bibnamefont{Wittlich}},
  \bibinfo{journal}{Nucl.Phys.} \textbf{\bibinfo{volume}{B457}},
  \bibinfo{pages}{577} (\bibinfo{year}{1995}), \eprint{9501001}.

\bibitem[{\citenamefont{van Baal}(1982)}]{vanBaal1982}
\bibinfo{author}{\bibfnamefont{P.}~\bibnamefont{van Baal}},
  \bibinfo{journal}{Commun. Math. Phys.} \textbf{\bibinfo{volume}{85}}
  (\bibinfo{year}{1982}), ISSN \bibinfo{issn}{0010-3616}.

\end{thebibliography}

\end{document}